\documentclass[onecolumn,unpublished,a4paper]{quantumarticle}
\usepackage[utf8]{inputenc}
\usepackage[backend=biber,style=alphabetic,maxbibnames=999]{biblatex}
\usepackage{amsmath}
\usepackage{amssymb}
\usepackage{amsthm}
\usepackage{amsfonts}
\usepackage[caption=false]{subfig}
\usepackage[colorlinks]{hyperref}
\usepackage[all]{hypcap}
\usepackage{tikz}
\usepackage{relsize}
\usepackage{color,soul}
\usepackage{capt-of}
\usepackage{mathtools}
\usepackage{float}
\usepackage[section]{placeins}
\usepackage{listings}
\usepackage[T1]{fontenc}  
\usetikzlibrary{decorations.pathreplacing}
\usepackage{braket}

\DeclareFixedFont{\ttb}{T1}{txtt}{bx}{n}{4}
\DeclareFixedFont{\ttm}{T1}{txtt}{m}{n}{4}
\definecolor{deepblue}{rgb}{0,0,0.5}
\definecolor{deepred}{rgb}{0.6,0,0}
\definecolor{deepgreen}{rgb}{0,0.5,0}
\newcommand\cppstyle{\lstset{
language=C++,
basicstyle=\ttm,
otherkeywords={uint8_t, __m256i, size_t, ASSERT_TRUE, EXPECT_TRUE, TEST, BENCHMARK},
keywordstyle=\ttb\color{deepblue},
emphstyle=\ttb\color{deepblue},
stringstyle=\color{deepgreen},
commentstyle=\fontfamily{txtt}\selectfont\color{gray},
showstringspaces=false,
literate={*}{{\char42}}1
         {-}{{\char45}}1
}}
\lstnewenvironment{cpp}[1][]
{\cppstyle\lstset{#1}}{}

\newcommand\pythonstyle{\lstset{
language=python,
basicstyle=\ttm,
morekeywords={assert,as,echo},
keywordstyle=\ttb\color{deepblue},
emphstyle=\ttb\color{deepblue},
stringstyle=\color{deepgreen},
commentstyle=\fontfamily{txtt}\selectfont\color{gray},
showstringspaces=false,
literate={*}{{\char42}}1
         {-}{{\char45}}1
}}
\lstnewenvironment{python}[1][]
{\pythonstyle\lstset{#1}}{}

\lstdefinestyle{stimcircuit}{
    language=python,
    basicstyle=\fontsize{6}{6}\selectfont\ttfamily,
    upquote=true,
    stepnumber=1,
    numbersep=8pt,
    showstringspaces=false,
    breaklines=true,
    frame=single,
    aboveskip=1.5em,
    belowskip=1.5em,
    commentstyle=\color{gray},
    classoffset=1,
    morekeywords={DETECTOR,OBSERVABLE_INCLUDE,rec},
    keywordstyle=\color{deepgreen},
    classoffset=2,
    morekeywords={H,R,MPP,M,RX,RY,MY,MX,SQRT\_X,XCY,XCZ,YCX},
    keywordstyle=\color{deepblue},
    classoffset=3,
    morekeywords={X_ERROR,DEPOLARIZE2,DEPOLARIZE1},
    keywordstyle=\color{red},
    classoffset=4,
    morekeywords={TICK,SHIFT_COORDS,QUBIT_COORDS},
    keywordstyle=\color{gray}
}

\hypersetup{linkcolor=blue}

\renewcommand\thesection{\arabic{section}}

\theoremstyle{definition}

\theoremstyle{definition}

\theoremstyle{definition}

\newcommand{\eq}[1]{\hyperref[eq:#1]{Equation~\ref*{eq:#1}}}
\renewcommand{\sec}[1]{\hyperref[sec:#1]{Section~\ref*{sec:#1}}}
\DeclareRobustCommand{\app}[1]{\hyperref[app:#1]{Appendix~\ref*{app:#1}}}
\newcommand{\fig}[1]{\hyperref[fig:#1]{Figure~\ref*{fig:#1}}}
\newcommand{\tab}[1]{\hyperref[tab:#1]{Table~\ref*{tab:#1}}}
\newcommand{\theoremref}[1]{\hyperref[theorem:#1]{Theorem~\ref*{theorem:#1}}}
\newcommand{\definitionref}[1]{\hyperref[definition:#1]{Definition~\ref*{definition:#1}}}

\begin{document}
\title{Yoked surface codes}

\date{\today}

\author{Craig Gidney}
\email{craig.gidney@gmail.com}
\affiliation{Google Quantum AI, Santa Barbara, California 93117, USA}

\author{Michael Newman}
\email{mgnewman@google.com}
\affiliation{Google Quantum AI, Santa Barbara, California 93117, USA}

\author{Peter Brooks}
\affiliation{Google, Sunnyvale, California 94089, USA}

\author{Cody Jones}
\affiliation{Google Quantum AI, Santa Barbara, California 93117, USA}

\begin{abstract}
We nearly triple the number of logical qubits per physical qubit of surface codes in the teraquop regime by concatenating them into high-density parity check codes.
These \emph{yoked surface codes} are arrayed in a rectangular grid, with parity checks (yokes) measured along each row, and optionally along each column, using lattice surgery.
Our construction assumes no additional connectivity beyond a nearest neighbor square qubit grid operating at a physical error rate of $10^{-3}$.

\end{abstract}

\maketitle

\tableofcontents

\section{Introduction}
\label{sec:introduction}

The surface code is a leading quantum error correcting code for building large scale fault-tolerant quantum computers because of its forgiving qubit quality and connectivity requirements~\cite{fowler2012surfacecodereview}.
The surface code's major downside is its extremely demanding quantity requirements.
At an error rate of $10^{-3}$, it takes 1000 to 2000 physical qubits per logical qubit for the surface code to reach error rates low enough to run classically intractable algorithms~\cite{shor1994,fowler2012surfacecodereview,gidney2021factor,soeken2020improved,litinskyhypercubeshor2023}.

There are many ideas in the field for reducing this overhead~\cite{tremblay2022ldpc,higgott2021hyperbolic,bravyi2011subsystembacon, breuckmann2017hyperbolic, berthusen2023partialmeasure, fawzi2020constant, li2020numerical, gottesman2013fault, kovalev2013fault, tillich2013quantum, panteleev2021degenerate, panteleev2022asymptotically, bravyi2023high, xu2023constant}, as well as bounds on possible improvements~\cite{bravyi2009no, bravyi2010tradeoffs, baspin2022quantifying, baspin2022connectivity, baspin2023improved}.
Constructions for reducing overhead frequently require high-fidelity long-range connections, which can be difficult to engineer in architectures like superconducting qubits.
When restricting to nearest neighbor planar connectivity, one strategy is to concatenate the surface code into an outer code with a higher ratio of logical qubits to physical qubits \cite{pattison2023planarcircuitbounds}.
The surface code provides high quality qubits, which the outer code can densely encode with increased protection, hopefully using fewer qubits than simply expanding the surface codes directly.
The surface code also provides mechanisms like lattice surgery~\cite{fowler2018latticesurgery,litinski2019gameofsurfacecodes} to perform operations between distant qubits. 

We usually imagine that the overlying code should have a high code distance, a high code rate, and low-density parity checks.
Small parity checks provide two important advantages.
First, their syndrome extraction circuits are small and highly parallelizable \cite{tremblay2022ldpc}, so the noise injected into the system while measuring checks is low.
Second, their locality limits the damage caused by correlated errors, sometimes for free \cite{manes2023distance}.
However, requiring these properties together can sometimes demand complex layouts and larger code block sizes to see improved performance.

\begin{figure}[htb!]
    \centering
    \resizebox{\linewidth}{!}{
     \includegraphics{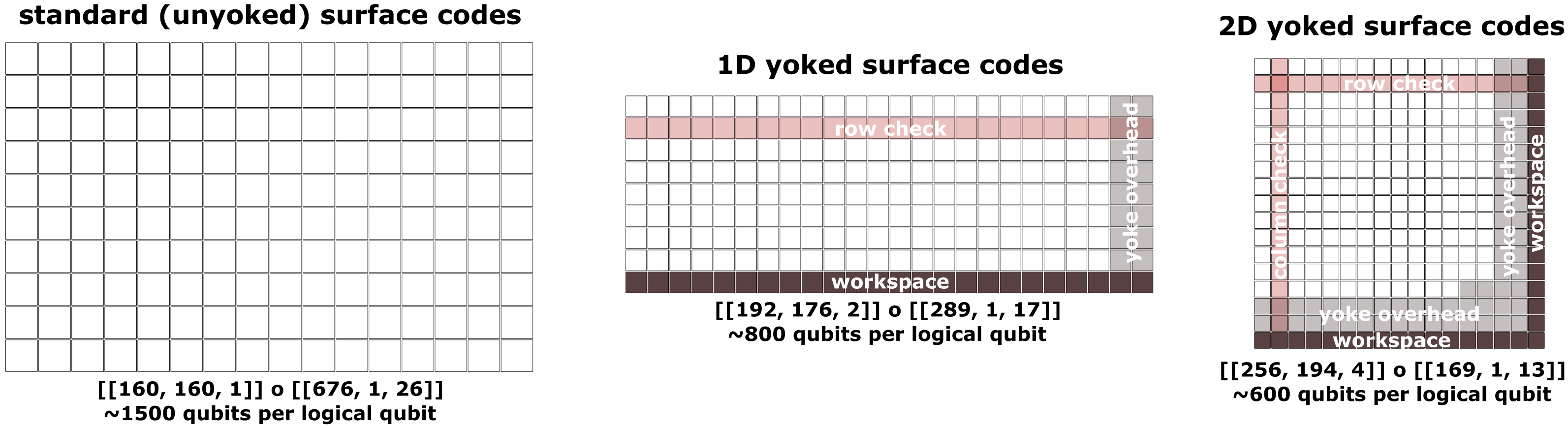}
    }
    \caption{
        From left to right: unyoked, 1D, and 2D yoked surface code patches.   
        In each row of 1D yoked surface codes, we measure multi-body logical $X$- and $Z$-type stabilizers.
        In 2D yoked surface codes, we additionally measure multi-body logical $X$- and $Z$-type stabilizers in each column.
        The $Z$-type stabilizers are applied to a permutation of the 2D code to commute with the $X$-type stabilizers.
        Grey patches represent overhead introduced by the stabilizers (i.e. ``yokes'').
        Dark patches represent the workspace required to measure the row/column stabilizers.
        There is also overhead due to interstitial space between patches for lattice surgery. 
        Concatenated code parameters, along with approximate overall qubit footprints (including the various overheads) labeled below.
        Note that the $[[192, 176, 2]]$ outer code is a collection of eight $[[24, 22, 2]]$ 1D parity check code blocks.
        All logical qubits can be reliably stored for about a trillion operations assuming a physical error rate of $10^{-3}$.
        The relative savings of yoked surface codes over unyoked surface codes grows as the target error rate decreases.
    }
    \label{fig:yoking_hierarchy}
\end{figure}

In this paper, we use simple \emph{parity check codes} as outer codes, focusing solely on achieving a high coding rate and simple layout, see \fig{yoking_hierarchy}.
We refer to these outer parity checks as \emph{yokes}.
In 1D, these consist of parity checks along each row of an array of surface codes \cite{steane1996simpleqec}.
In 2D, they consist of parity checks along each row and column of the array, up to qubit permutations.
These outer codes have distances two and four respectively which, by utilizing the soft information provided by the inner surface codes \cite{poulin2006optimal}, double and quadruple the inner code distance.
The inner surface codes suppress error rates to levels that allow us to measure high-weight checks without incurring significant noise.
Furthermore, we can avoid damaging correlated errors introduced when measuring high-weight checks by adding protection against them using a spacetime tradeoff during lattice surgery.

Simulations of yoked surface codes can grow quite expensive.
In the largest cases we consider, we may want to compute millions of shots of hundreds of surface code patches over thousands of rounds.
Consequently, we perform two types of simplified simulations to estimate the overhead of yoked surface codes.
The first is a smaller, full circuit simulation of the inner codes concatenated into a one-round simulation of the outer code.
Specifically, we simulate many patches of surface codes with error graphs connected by yoke detectors running along their boundaries.
To remain tractable, these simulations are limited to hundreds of rounds of inner surface code cycles.

In order to test sizes that are representative of the full spacetime volume required to measure several outer code cycles, we also run multi-round phenomenological simulations of the outer code while replacing the full simulation of the inner code with a method we call \emph{gap simulation}.
These are based on sampling \emph{complementary gaps} \cite{hutter2014complementarygap, bombin2022postselect}, which are the log-likelihood ratios of the probabilities of minimum-weight matchings produced by a (correlated) minimum-weight perfect matching decoder when constrained to flip/not flip a logical observable.
Gap simulation directly samples from the distribution of complementary gaps in the surface code to obtain both a success/failure of the minimum-weight matching to identify the correct error, as well as a confidence in that prediction.
We compute an empirical distribution for complementary gaps of surface codes over multiple distances, and use a simple heuristic to extrapolate these gap distributions to many rounds.
This allows us to sample gaps directly, reducing many rounds of surface code simulation to a single sampled probability of failure. 
We validate the gap simulations by comparing them to the smaller circuit-level simulations of the inner code with a single outer round.
We then use gap sampling to run larger phenomenological simulations of the outer code, which we expect to be representative of the full circuit-level performance.

There are different choices of inner code distance, block size, and number of blocks that provide the smallest overhead for achieving different target logical error rates.
Finding simple heuristics for the logical error rate in terms of these parameters is then helpful for identifying the best layout for a particular target.
If the error rate is low enough, then the dominant error mechanisms should be near-minimum-weight failure paths.
Consequently, we approximate the logical error rates using simple path-counting heuristics, and observe reasonable agreement between the error rates predicted from simulation and the error rates predicted from the path-counting formulas.
Finally, we use these heuristic formulas to estimate the overhead of yoked surface codes, identifying the most compact layouts that achieve different target logical error rates.

The paper proceeds as follows.
In \sec{parity_check_codes}, we describe the 1D and 2D quantum parity check codes that form the outer codes of our concatenated schemes.
In \sec{lattice_surgery}, we provide lattice surgery constructions for measuring these checks, and analyze their overhead and fault-tolerance properties.
In \sec{complementary_gaps}, we provide empirical estimates of complementary gap distributions of the surface code, and extrapolate these distributions to many rounds for use in gap simulation.
Finally, in \sec{benchmarking}, we present benchmarks from Monte Carlo simulations of yoked surface codes to estimate their qubit savings compared to standard surface codes.
In \sec{conclusion} we make some closing remarks.
The paper also contains \app{noise}, which specifies the circuit noise model we use in simulations, \app{y-type} which discusses an alternative outer code construction using a single $Y$-type check, and \app{mdpc} which describes higher-dimensional generalizations of quantum parity check codes.

\section{Quantum parity check codes}
\label{sec:parity_check_codes}

Quantum parity check codes are CSS codes that generalize classical parity check (or ``array'') codes. 
In classical parity check codes, bits are laid out in a 1D or 2D array, and the parity of each row or each row and column are checked.
We focus on these codes because their parity checks are geometrically simple, and their rate quickly tends to one as their code block size increases.

Unlike classical parity check codes, we must enforce certain block size restrictions to ensure that the stabilizers of quantum parity check codes commute.
In 1D, we require the length of the code to be divisible by two.
In 2D, we require the side length $n$ of the block to be divisible by four.
The reason is that row and column operators of opposite Pauli type anti-commute, since they intersect in a single location.
When $4 \mid n$, this can be fixed by permuting the qubits for different Pauli type row and column checks.
Let $e_i \in \mathbb{F}_2^{n/2} \times \mathbb{F}_2^{n/2}$ and $f_j \in \mathbb{F}_2^{2} \times \mathbb{F}_2^{2}$, where $(e_i)_{kl} = \delta_{ik}$ and $(f_j)_{kl} = \delta_{jk}$. 
Then we can modify the support of the $Z$-type row and column parity checks of an $n\times n$ 2D parity check code as
\begin{align*}
e_i \otimes f_j &\mapsto e_i \otimes f_j^\top \\
e_i^\top \otimes f_j^\top &\mapsto e_i^\top \otimes f_j
\end{align*}
to ensure that they commute with row and column $X$-type checks.
As an example, we describe the stabilizers and observables of the $[[64,34,4]]$ 2D parity check code in \tab{squareberg_code}. 

\begin{table}[htb!]
    \centering
    \resizebox{\linewidth}{!}{
        \begin{tabular}{|c|c|c|c||c|c|c|c|c|c|c|c|}
\hline
\multicolumn{12}{|c|}{[[64, 34, 4]] 2D Parity Check Code}
\\\hline
{\small $X$ Col Checks} & {\small $X$ Row Checks} & {\small $Z$ Bi-Col Checks} & {\small $Z$ Bi-Row Checks} & \multicolumn{8}{|c|}{$Y$ Observables (positioned by location of $Y$ term)}
\\\hline
\begin{tabular}{c}
    \texttt{X\_\_\_\_\_\_\_}\\
    \texttt{X\_\_\_\_\_\_\_}\\
    \texttt{X\_\_\_\_\_\_\_}\\
    \texttt{X\_\_\_\_\_\_\_}\\
    \texttt{X\_\_\_\_\_\_\_}\\
    \texttt{X\_\_\_\_\_\_\_}\\
    \texttt{X\_\_\_\_\_\_\_}\\
    \texttt{X\_\_\_\_\_\_\_}\\
\end{tabular}&\begin{tabular}{c}
    \texttt{XXXXXXXX}\\
    \texttt{\_\_\_\_\_\_\_\_}\\
    \texttt{\_\_\_\_\_\_\_\_}\\
    \texttt{\_\_\_\_\_\_\_\_}\\
    \texttt{\_\_\_\_\_\_\_\_}\\
    \texttt{\_\_\_\_\_\_\_\_}\\
    \texttt{\_\_\_\_\_\_\_\_}\\
    \texttt{\_\_\_\_\_\_\_\_}\\
\end{tabular}&\begin{tabular}{c}
    \texttt{ZZ\_\_\_\_\_\_}\\
    \texttt{\_\_\_\_\_\_\_\_}\\
    \texttt{ZZ\_\_\_\_\_\_}\\
    \texttt{\_\_\_\_\_\_\_\_}\\
    \texttt{ZZ\_\_\_\_\_\_}\\
    \texttt{\_\_\_\_\_\_\_\_}\\
    \texttt{ZZ\_\_\_\_\_\_}\\
    \texttt{\_\_\_\_\_\_\_\_}\\
\end{tabular}&\begin{tabular}{c}
    \texttt{Z\_Z\_Z\_Z\_}\\
    \texttt{Z\_Z\_Z\_Z\_}\\
    \texttt{\_\_\_\_\_\_\_\_}\\
    \texttt{\_\_\_\_\_\_\_\_}\\
    \texttt{\_\_\_\_\_\_\_\_}\\
    \texttt{\_\_\_\_\_\_\_\_}\\
    \texttt{\_\_\_\_\_\_\_\_}\\
    \texttt{\_\_\_\_\_\_\_\_}\\
\end{tabular}&\texttt{{\kern 5em}}&\texttt{{\kern 5em}}&\texttt{{\kern 5em}}&\texttt{{\kern 5em}}&\texttt{{\kern 5em}}&\texttt{{\kern 5em}}&\texttt{{\kern 5em}}&\texttt{{\kern 5em}}\\\hline
\begin{tabular}{c}
    \texttt{\_X\_\_\_\_\_\_}\\
    \texttt{\_X\_\_\_\_\_\_}\\
    \texttt{\_X\_\_\_\_\_\_}\\
    \texttt{\_X\_\_\_\_\_\_}\\
    \texttt{\_X\_\_\_\_\_\_}\\
    \texttt{\_X\_\_\_\_\_\_}\\
    \texttt{\_X\_\_\_\_\_\_}\\
    \texttt{\_X\_\_\_\_\_\_}\\
\end{tabular}&\begin{tabular}{c}
    \texttt{\_\_\_\_\_\_\_\_}\\
    \texttt{XXXXXXXX}\\
    \texttt{\_\_\_\_\_\_\_\_}\\
    \texttt{\_\_\_\_\_\_\_\_}\\
    \texttt{\_\_\_\_\_\_\_\_}\\
    \texttt{\_\_\_\_\_\_\_\_}\\
    \texttt{\_\_\_\_\_\_\_\_}\\
    \texttt{\_\_\_\_\_\_\_\_}\\
\end{tabular}&\begin{tabular}{c}
    \texttt{\_\_ZZ\_\_\_\_}\\
    \texttt{\_\_\_\_\_\_\_\_}\\
    \texttt{\_\_ZZ\_\_\_\_}\\
    \texttt{\_\_\_\_\_\_\_\_}\\
    \texttt{\_\_ZZ\_\_\_\_}\\
    \texttt{\_\_\_\_\_\_\_\_}\\
    \texttt{\_\_ZZ\_\_\_\_}\\
    \texttt{\_\_\_\_\_\_\_\_}\\
\end{tabular}&\begin{tabular}{c}
    \texttt{\_Z\_Z\_Z\_Z}\\
    \texttt{\_Z\_Z\_Z\_Z}\\
    \texttt{\_\_\_\_\_\_\_\_}\\
    \texttt{\_\_\_\_\_\_\_\_}\\
    \texttt{\_\_\_\_\_\_\_\_}\\
    \texttt{\_\_\_\_\_\_\_\_}\\
    \texttt{\_\_\_\_\_\_\_\_}\\
    \texttt{\_\_\_\_\_\_\_\_}\\
\end{tabular}&\texttt{{\kern 5em}}&\begin{tabular}{c}
    \texttt{ZZ\_\_\_\_\_\_}\\
    \texttt{ZY\_\_\_\_\_X}\\
    \texttt{\_\_\_\_\_\_\_\_}\\
    \texttt{\_\_\_\_\_\_\_\_}\\
    \texttt{\_\_\_\_\_\_\_\_}\\
    \texttt{\_\_\_\_\_\_\_\_}\\
    \texttt{\_\_\_\_\_\_\_\_}\\
    \texttt{\_X\_\_\_\_\_X}\\
\end{tabular}&\begin{tabular}{c}
    \texttt{Z\_Z\_\_\_\_\_}\\
    \texttt{Z\_Y\_\_\_X\_}\\
    \texttt{\_\_\_\_\_\_\_\_}\\
    \texttt{\_\_\_\_\_\_\_\_}\\
    \texttt{\_\_\_\_\_\_\_\_}\\
    \texttt{\_\_\_\_\_\_\_\_}\\
    \texttt{\_\_\_\_\_\_\_\_}\\
    \texttt{\_\_\_X\_\_\_X}\\
\end{tabular}&\begin{tabular}{c}
    \texttt{Z\_\_Z\_\_\_\_}\\
    \texttt{Z\_\_Y\_\_\_X}\\
    \texttt{\_\_\_\_\_\_\_\_}\\
    \texttt{\_\_\_\_\_\_\_\_}\\
    \texttt{\_\_\_\_\_\_\_\_}\\
    \texttt{\_\_\_\_\_\_\_\_}\\
    \texttt{\_\_\_\_\_\_\_\_}\\
    \texttt{\_\_\_X\_\_\_X}\\
\end{tabular}&\begin{tabular}{c}
    \texttt{Z\_\_\_Z\_\_\_}\\
    \texttt{Z\_\_\_Y\_X\_}\\
    \texttt{\_\_\_\_\_\_\_\_}\\
    \texttt{\_\_\_\_\_\_\_\_}\\
    \texttt{\_\_\_\_\_\_\_\_}\\
    \texttt{\_\_\_\_\_\_\_\_}\\
    \texttt{\_\_\_\_\_\_\_\_}\\
    \texttt{\_\_\_\_\_X\_X}\\
\end{tabular}&\begin{tabular}{c}
    \texttt{Z\_\_\_\_Z\_\_}\\
    \texttt{Z\_\_\_\_Y\_X}\\
    \texttt{\_\_\_\_\_\_\_\_}\\
    \texttt{\_\_\_\_\_\_\_\_}\\
    \texttt{\_\_\_\_\_\_\_\_}\\
    \texttt{\_\_\_\_\_\_\_\_}\\
    \texttt{\_\_\_\_\_\_\_\_}\\
    \texttt{\_\_\_\_\_X\_X}\\
\end{tabular}&\texttt{{\kern 5em}}&\texttt{{\kern 5em}}\\\hline
\begin{tabular}{c}
    \texttt{\_\_X\_\_\_\_\_}\\
    \texttt{\_\_X\_\_\_\_\_}\\
    \texttt{\_\_X\_\_\_\_\_}\\
    \texttt{\_\_X\_\_\_\_\_}\\
    \texttt{\_\_X\_\_\_\_\_}\\
    \texttt{\_\_X\_\_\_\_\_}\\
    \texttt{\_\_X\_\_\_\_\_}\\
    \texttt{\_\_X\_\_\_\_\_}\\
\end{tabular}&\begin{tabular}{c}
    \texttt{\_\_\_\_\_\_\_\_}\\
    \texttt{\_\_\_\_\_\_\_\_}\\
    \texttt{XXXXXXXX}\\
    \texttt{\_\_\_\_\_\_\_\_}\\
    \texttt{\_\_\_\_\_\_\_\_}\\
    \texttt{\_\_\_\_\_\_\_\_}\\
    \texttt{\_\_\_\_\_\_\_\_}\\
    \texttt{\_\_\_\_\_\_\_\_}\\
\end{tabular}&\begin{tabular}{c}
    \texttt{\_\_\_\_ZZ\_\_}\\
    \texttt{\_\_\_\_\_\_\_\_}\\
    \texttt{\_\_\_\_ZZ\_\_}\\
    \texttt{\_\_\_\_\_\_\_\_}\\
    \texttt{\_\_\_\_ZZ\_\_}\\
    \texttt{\_\_\_\_\_\_\_\_}\\
    \texttt{\_\_\_\_ZZ\_\_}\\
    \texttt{\_\_\_\_\_\_\_\_}\\
\end{tabular}&\begin{tabular}{c}
    \texttt{\_\_\_\_\_\_\_\_}\\
    \texttt{\_\_\_\_\_\_\_\_}\\
    \texttt{Z\_Z\_Z\_Z\_}\\
    \texttt{Z\_Z\_Z\_Z\_}\\
    \texttt{\_\_\_\_\_\_\_\_}\\
    \texttt{\_\_\_\_\_\_\_\_}\\
    \texttt{\_\_\_\_\_\_\_\_}\\
    \texttt{\_\_\_\_\_\_\_\_}\\
\end{tabular}&\texttt{{\kern 5em}}&\begin{tabular}{c}
    \texttt{ZZ\_\_\_\_\_\_}\\
    \texttt{\_\_\_\_\_\_\_\_}\\
    \texttt{ZY\_\_\_\_\_\_}\\
    \texttt{\_\_\_\_\_\_\_X}\\
    \texttt{\_\_\_\_\_\_\_\_}\\
    \texttt{\_\_\_\_\_\_\_\_}\\
    \texttt{\_X\_\_\_\_\_\_}\\
    \texttt{\_\_\_\_\_\_\_X}\\
\end{tabular}&\begin{tabular}{c}
    \texttt{Z\_Z\_\_\_\_\_}\\
    \texttt{\_\_\_\_\_\_\_\_}\\
    \texttt{Z\_Y\_\_\_\_\_}\\
    \texttt{\_\_\_\_\_\_X\_}\\
    \texttt{\_\_\_\_\_\_\_\_}\\
    \texttt{\_\_\_\_\_\_\_\_}\\
    \texttt{\_\_\_X\_\_\_\_}\\
    \texttt{\_\_\_\_\_\_\_X}\\
\end{tabular}&\begin{tabular}{c}
    \texttt{Z\_\_Z\_\_\_\_}\\
    \texttt{\_\_\_\_\_\_\_\_}\\
    \texttt{Z\_\_Y\_\_\_\_}\\
    \texttt{\_\_\_\_\_\_\_X}\\
    \texttt{\_\_\_\_\_\_\_\_}\\
    \texttt{\_\_\_\_\_\_\_\_}\\
    \texttt{\_\_\_X\_\_\_\_}\\
    \texttt{\_\_\_\_\_\_\_X}\\
\end{tabular}&\begin{tabular}{c}
    \texttt{Z\_\_\_Z\_\_\_}\\
    \texttt{\_\_\_\_\_\_\_\_}\\
    \texttt{Z\_\_\_Y\_\_\_}\\
    \texttt{\_\_\_\_\_\_X\_}\\
    \texttt{\_\_\_\_\_\_\_\_}\\
    \texttt{\_\_\_\_\_\_\_\_}\\
    \texttt{\_\_\_\_\_X\_\_}\\
    \texttt{\_\_\_\_\_\_\_X}\\
\end{tabular}&\begin{tabular}{c}
    \texttt{Z\_\_\_\_Z\_\_}\\
    \texttt{\_\_\_\_\_\_\_\_}\\
    \texttt{Z\_\_\_\_Y\_\_}\\
    \texttt{\_\_\_\_\_\_\_X}\\
    \texttt{\_\_\_\_\_\_\_\_}\\
    \texttt{\_\_\_\_\_\_\_\_}\\
    \texttt{\_\_\_\_\_X\_\_}\\
    \texttt{\_\_\_\_\_\_\_X}\\
\end{tabular}&\begin{tabular}{c}
    \texttt{Z\_\_\_\_\_Z\_}\\
    \texttt{\_\_\_\_\_\_\_\_}\\
    \texttt{Z\_\_\_\_\_Y\_}\\
    \texttt{\_\_\_\_\_\_X\_}\\
    \texttt{\_\_\_\_\_\_\_\_}\\
    \texttt{\_\_\_\_\_\_\_\_}\\
    \texttt{\_\_\_\_\_\_\_X}\\
    \texttt{\_\_\_\_\_\_\_X}\\
\end{tabular}&\begin{tabular}{c}
    \texttt{Z\_\_\_\_\_\_Z}\\
    \texttt{\_\_\_\_\_\_\_\_}\\
    \texttt{Z\_\_\_\_\_\_Y}\\
    \texttt{\_\_\_\_\_\_\_X}\\
    \texttt{\_\_\_\_\_\_\_\_}\\
    \texttt{\_\_\_\_\_\_\_\_}\\
    \texttt{\_\_\_\_\_\_\_X}\\
    \texttt{\_\_\_\_\_\_\_X}\\
\end{tabular}\\\hline
\begin{tabular}{c}
    \texttt{\_\_\_X\_\_\_\_}\\
    \texttt{\_\_\_X\_\_\_\_}\\
    \texttt{\_\_\_X\_\_\_\_}\\
    \texttt{\_\_\_X\_\_\_\_}\\
    \texttt{\_\_\_X\_\_\_\_}\\
    \texttt{\_\_\_X\_\_\_\_}\\
    \texttt{\_\_\_X\_\_\_\_}\\
    \texttt{\_\_\_X\_\_\_\_}\\
\end{tabular}&\begin{tabular}{c}
    \texttt{\_\_\_\_\_\_\_\_}\\
    \texttt{\_\_\_\_\_\_\_\_}\\
    \texttt{\_\_\_\_\_\_\_\_}\\
    \texttt{XXXXXXXX}\\
    \texttt{\_\_\_\_\_\_\_\_}\\
    \texttt{\_\_\_\_\_\_\_\_}\\
    \texttt{\_\_\_\_\_\_\_\_}\\
    \texttt{\_\_\_\_\_\_\_\_}\\
\end{tabular}&\begin{tabular}{c}
    \texttt{\_\_\_\_\_\_ZZ}\\
    \texttt{\_\_\_\_\_\_\_\_}\\
    \texttt{\_\_\_\_\_\_ZZ}\\
    \texttt{\_\_\_\_\_\_\_\_}\\
    \texttt{\_\_\_\_\_\_ZZ}\\
    \texttt{\_\_\_\_\_\_\_\_}\\
    \texttt{\_\_\_\_\_\_ZZ}\\
    \texttt{\_\_\_\_\_\_\_\_}\\
\end{tabular}&\begin{tabular}{c}
    \texttt{\_\_\_\_\_\_\_\_}\\
    \texttt{\_\_\_\_\_\_\_\_}\\
    \texttt{\_Z\_Z\_Z\_Z}\\
    \texttt{\_Z\_Z\_Z\_Z}\\
    \texttt{\_\_\_\_\_\_\_\_}\\
    \texttt{\_\_\_\_\_\_\_\_}\\
    \texttt{\_\_\_\_\_\_\_\_}\\
    \texttt{\_\_\_\_\_\_\_\_}\\
\end{tabular}&\texttt{{\kern 5em}}&\begin{tabular}{c}
    \texttt{ZZ\_\_\_\_\_\_}\\
    \texttt{\_\_\_\_\_\_\_\_}\\
    \texttt{\_\_\_\_\_\_\_\_}\\
    \texttt{ZY\_\_\_\_\_X}\\
    \texttt{\_\_\_\_\_\_\_\_}\\
    \texttt{\_\_\_\_\_\_\_\_}\\
    \texttt{\_\_\_\_\_\_\_\_}\\
    \texttt{\_X\_\_\_\_\_X}\\
\end{tabular}&\begin{tabular}{c}
    \texttt{Z\_Z\_\_\_\_\_}\\
    \texttt{\_\_\_\_\_\_\_\_}\\
    \texttt{\_\_\_\_\_\_\_\_}\\
    \texttt{Z\_Y\_\_\_X\_}\\
    \texttt{\_\_\_\_\_\_\_\_}\\
    \texttt{\_\_\_\_\_\_\_\_}\\
    \texttt{\_\_\_\_\_\_\_\_}\\
    \texttt{\_\_\_X\_\_\_X}\\
\end{tabular}&\begin{tabular}{c}
    \texttt{Z\_\_Z\_\_\_\_}\\
    \texttt{\_\_\_\_\_\_\_\_}\\
    \texttt{\_\_\_\_\_\_\_\_}\\
    \texttt{Z\_\_Y\_\_\_X}\\
    \texttt{\_\_\_\_\_\_\_\_}\\
    \texttt{\_\_\_\_\_\_\_\_}\\
    \texttt{\_\_\_\_\_\_\_\_}\\
    \texttt{\_\_\_X\_\_\_X}\\
\end{tabular}&\begin{tabular}{c}
    \texttt{Z\_\_\_Z\_\_\_}\\
    \texttt{\_\_\_\_\_\_\_\_}\\
    \texttt{\_\_\_\_\_\_\_\_}\\
    \texttt{Z\_\_\_Y\_X\_}\\
    \texttt{\_\_\_\_\_\_\_\_}\\
    \texttt{\_\_\_\_\_\_\_\_}\\
    \texttt{\_\_\_\_\_\_\_\_}\\
    \texttt{\_\_\_\_\_X\_X}\\
\end{tabular}&\begin{tabular}{c}
    \texttt{Z\_\_\_\_Z\_\_}\\
    \texttt{\_\_\_\_\_\_\_\_}\\
    \texttt{\_\_\_\_\_\_\_\_}\\
    \texttt{Z\_\_\_\_Y\_X}\\
    \texttt{\_\_\_\_\_\_\_\_}\\
    \texttt{\_\_\_\_\_\_\_\_}\\
    \texttt{\_\_\_\_\_\_\_\_}\\
    \texttt{\_\_\_\_\_X\_X}\\
\end{tabular}&\texttt{{\kern 5em}}&\texttt{{\kern 5em}}\\\hline
\begin{tabular}{c}
    \texttt{\_\_\_\_X\_\_\_}\\
    \texttt{\_\_\_\_X\_\_\_}\\
    \texttt{\_\_\_\_X\_\_\_}\\
    \texttt{\_\_\_\_X\_\_\_}\\
    \texttt{\_\_\_\_X\_\_\_}\\
    \texttt{\_\_\_\_X\_\_\_}\\
    \texttt{\_\_\_\_X\_\_\_}\\
    \texttt{\_\_\_\_X\_\_\_}\\
\end{tabular}&\begin{tabular}{c}
    \texttt{\_\_\_\_\_\_\_\_}\\
    \texttt{\_\_\_\_\_\_\_\_}\\
    \texttt{\_\_\_\_\_\_\_\_}\\
    \texttt{\_\_\_\_\_\_\_\_}\\
    \texttt{XXXXXXXX}\\
    \texttt{\_\_\_\_\_\_\_\_}\\
    \texttt{\_\_\_\_\_\_\_\_}\\
    \texttt{\_\_\_\_\_\_\_\_}\\
\end{tabular}&\begin{tabular}{c}
    \texttt{\_\_\_\_\_\_\_\_}\\
    \texttt{ZZ\_\_\_\_\_\_}\\
    \texttt{\_\_\_\_\_\_\_\_}\\
    \texttt{ZZ\_\_\_\_\_\_}\\
    \texttt{\_\_\_\_\_\_\_\_}\\
    \texttt{ZZ\_\_\_\_\_\_}\\
    \texttt{\_\_\_\_\_\_\_\_}\\
    \texttt{ZZ\_\_\_\_\_\_}\\
\end{tabular}&\begin{tabular}{c}
    \texttt{\_\_\_\_\_\_\_\_}\\
    \texttt{\_\_\_\_\_\_\_\_}\\
    \texttt{\_\_\_\_\_\_\_\_}\\
    \texttt{\_\_\_\_\_\_\_\_}\\
    \texttt{Z\_Z\_Z\_Z\_}\\
    \texttt{Z\_Z\_Z\_Z\_}\\
    \texttt{\_\_\_\_\_\_\_\_}\\
    \texttt{\_\_\_\_\_\_\_\_}\\
\end{tabular}&\texttt{{\kern 5em}}&\begin{tabular}{c}
    \texttt{ZZ\_\_\_\_\_\_}\\
    \texttt{\_\_\_\_\_\_\_\_}\\
    \texttt{\_\_\_\_\_\_\_\_}\\
    \texttt{\_\_\_\_\_\_\_\_}\\
    \texttt{ZY\_\_\_\_\_\_}\\
    \texttt{\_\_\_\_\_\_\_X}\\
    \texttt{\_X\_\_\_\_\_\_}\\
    \texttt{\_\_\_\_\_\_\_X}\\
\end{tabular}&\begin{tabular}{c}
    \texttt{Z\_Z\_\_\_\_\_}\\
    \texttt{\_\_\_\_\_\_\_\_}\\
    \texttt{\_\_\_\_\_\_\_\_}\\
    \texttt{\_\_\_\_\_\_\_\_}\\
    \texttt{Z\_Y\_\_\_\_\_}\\
    \texttt{\_\_\_\_\_\_X\_}\\
    \texttt{\_\_\_X\_\_\_\_}\\
    \texttt{\_\_\_\_\_\_\_X}\\
\end{tabular}&\begin{tabular}{c}
    \texttt{Z\_\_Z\_\_\_\_}\\
    \texttt{\_\_\_\_\_\_\_\_}\\
    \texttt{\_\_\_\_\_\_\_\_}\\
    \texttt{\_\_\_\_\_\_\_\_}\\
    \texttt{Z\_\_Y\_\_\_\_}\\
    \texttt{\_\_\_\_\_\_\_X}\\
    \texttt{\_\_\_X\_\_\_\_}\\
    \texttt{\_\_\_\_\_\_\_X}\\
\end{tabular}&\begin{tabular}{c}
    \texttt{Z\_\_\_Z\_\_\_}\\
    \texttt{\_\_\_\_\_\_\_\_}\\
    \texttt{\_\_\_\_\_\_\_\_}\\
    \texttt{\_\_\_\_\_\_\_\_}\\
    \texttt{Z\_\_\_Y\_\_\_}\\
    \texttt{\_\_\_\_\_\_X\_}\\
    \texttt{\_\_\_\_\_X\_\_}\\
    \texttt{\_\_\_\_\_\_\_X}\\
\end{tabular}&\begin{tabular}{c}
    \texttt{Z\_\_\_\_Z\_\_}\\
    \texttt{\_\_\_\_\_\_\_\_}\\
    \texttt{\_\_\_\_\_\_\_\_}\\
    \texttt{\_\_\_\_\_\_\_\_}\\
    \texttt{Z\_\_\_\_Y\_\_}\\
    \texttt{\_\_\_\_\_\_\_X}\\
    \texttt{\_\_\_\_\_X\_\_}\\
    \texttt{\_\_\_\_\_\_\_X}\\
\end{tabular}&\begin{tabular}{c}
    \texttt{Z\_\_\_\_\_Z\_}\\
    \texttt{\_\_\_\_\_\_\_\_}\\
    \texttt{\_\_\_\_\_\_\_\_}\\
    \texttt{\_\_\_\_\_\_\_\_}\\
    \texttt{Z\_\_\_\_\_Y\_}\\
    \texttt{\_\_\_\_\_\_X\_}\\
    \texttt{\_\_\_\_\_\_\_X}\\
    \texttt{\_\_\_\_\_\_\_X}\\
\end{tabular}&\begin{tabular}{c}
    \texttt{Z\_\_\_\_\_\_Z}\\
    \texttt{\_\_\_\_\_\_\_\_}\\
    \texttt{\_\_\_\_\_\_\_\_}\\
    \texttt{\_\_\_\_\_\_\_\_}\\
    \texttt{Z\_\_\_\_\_\_Y}\\
    \texttt{\_\_\_\_\_\_\_X}\\
    \texttt{\_\_\_\_\_\_\_X}\\
    \texttt{\_\_\_\_\_\_\_X}\\
\end{tabular}\\\hline
\begin{tabular}{c}
    \texttt{\_\_\_\_\_X\_\_}\\
    \texttt{\_\_\_\_\_X\_\_}\\
    \texttt{\_\_\_\_\_X\_\_}\\
    \texttt{\_\_\_\_\_X\_\_}\\
    \texttt{\_\_\_\_\_X\_\_}\\
    \texttt{\_\_\_\_\_X\_\_}\\
    \texttt{\_\_\_\_\_X\_\_}\\
    \texttt{\_\_\_\_\_X\_\_}\\
\end{tabular}&\begin{tabular}{c}
    \texttt{\_\_\_\_\_\_\_\_}\\
    \texttt{\_\_\_\_\_\_\_\_}\\
    \texttt{\_\_\_\_\_\_\_\_}\\
    \texttt{\_\_\_\_\_\_\_\_}\\
    \texttt{\_\_\_\_\_\_\_\_}\\
    \texttt{XXXXXXXX}\\
    \texttt{\_\_\_\_\_\_\_\_}\\
    \texttt{\_\_\_\_\_\_\_\_}\\
\end{tabular}&\begin{tabular}{c}
    \texttt{\_\_\_\_\_\_\_\_}\\
    \texttt{\_\_ZZ\_\_\_\_}\\
    \texttt{\_\_\_\_\_\_\_\_}\\
    \texttt{\_\_ZZ\_\_\_\_}\\
    \texttt{\_\_\_\_\_\_\_\_}\\
    \texttt{\_\_ZZ\_\_\_\_}\\
    \texttt{\_\_\_\_\_\_\_\_}\\
    \texttt{\_\_ZZ\_\_\_\_}\\
\end{tabular}&\begin{tabular}{c}
    \texttt{\_\_\_\_\_\_\_\_}\\
    \texttt{\_\_\_\_\_\_\_\_}\\
    \texttt{\_\_\_\_\_\_\_\_}\\
    \texttt{\_\_\_\_\_\_\_\_}\\
    \texttt{\_Z\_Z\_Z\_Z}\\
    \texttt{\_Z\_Z\_Z\_Z}\\
    \texttt{\_\_\_\_\_\_\_\_}\\
    \texttt{\_\_\_\_\_\_\_\_}\\
\end{tabular}&\texttt{{\kern 5em}}&\begin{tabular}{c}
    \texttt{ZZ\_\_\_\_\_\_}\\
    \texttt{\_\_\_\_\_\_\_\_}\\
    \texttt{\_\_\_\_\_\_\_\_}\\
    \texttt{\_\_\_\_\_\_\_\_}\\
    \texttt{\_\_\_\_\_\_\_\_}\\
    \texttt{ZY\_\_\_\_\_X}\\
    \texttt{\_\_\_\_\_\_\_\_}\\
    \texttt{\_X\_\_\_\_\_X}\\
\end{tabular}&\begin{tabular}{c}
    \texttt{Z\_Z\_\_\_\_\_}\\
    \texttt{\_\_\_\_\_\_\_\_}\\
    \texttt{\_\_\_\_\_\_\_\_}\\
    \texttt{\_\_\_\_\_\_\_\_}\\
    \texttt{\_\_\_\_\_\_\_\_}\\
    \texttt{Z\_Y\_\_\_X\_}\\
    \texttt{\_\_\_\_\_\_\_\_}\\
    \texttt{\_\_\_X\_\_\_X}\\
\end{tabular}&\begin{tabular}{c}
    \texttt{Z\_\_Z\_\_\_\_}\\
    \texttt{\_\_\_\_\_\_\_\_}\\
    \texttt{\_\_\_\_\_\_\_\_}\\
    \texttt{\_\_\_\_\_\_\_\_}\\
    \texttt{\_\_\_\_\_\_\_\_}\\
    \texttt{Z\_\_Y\_\_\_X}\\
    \texttt{\_\_\_\_\_\_\_\_}\\
    \texttt{\_\_\_X\_\_\_X}\\
\end{tabular}&\begin{tabular}{c}
    \texttt{Z\_\_\_Z\_\_\_}\\
    \texttt{\_\_\_\_\_\_\_\_}\\
    \texttt{\_\_\_\_\_\_\_\_}\\
    \texttt{\_\_\_\_\_\_\_\_}\\
    \texttt{\_\_\_\_\_\_\_\_}\\
    \texttt{Z\_\_\_Y\_X\_}\\
    \texttt{\_\_\_\_\_\_\_\_}\\
    \texttt{\_\_\_\_\_X\_X}\\
\end{tabular}&\begin{tabular}{c}
    \texttt{Z\_\_\_\_Z\_\_}\\
    \texttt{\_\_\_\_\_\_\_\_}\\
    \texttt{\_\_\_\_\_\_\_\_}\\
    \texttt{\_\_\_\_\_\_\_\_}\\
    \texttt{\_\_\_\_\_\_\_\_}\\
    \texttt{Z\_\_\_\_Y\_X}\\
    \texttt{\_\_\_\_\_\_\_\_}\\
    \texttt{\_\_\_\_\_X\_X}\\
\end{tabular}&\texttt{{\kern 5em}}&\texttt{{\kern 5em}}\\\hline
\begin{tabular}{c}
    \texttt{\_\_\_\_\_\_X\_}\\
    \texttt{\_\_\_\_\_\_X\_}\\
    \texttt{\_\_\_\_\_\_X\_}\\
    \texttt{\_\_\_\_\_\_X\_}\\
    \texttt{\_\_\_\_\_\_X\_}\\
    \texttt{\_\_\_\_\_\_X\_}\\
    \texttt{\_\_\_\_\_\_X\_}\\
    \texttt{\_\_\_\_\_\_X\_}\\
\end{tabular}&\begin{tabular}{c}
    \texttt{\_\_\_\_\_\_\_\_}\\
    \texttt{\_\_\_\_\_\_\_\_}\\
    \texttt{\_\_\_\_\_\_\_\_}\\
    \texttt{\_\_\_\_\_\_\_\_}\\
    \texttt{\_\_\_\_\_\_\_\_}\\
    \texttt{\_\_\_\_\_\_\_\_}\\
    \texttt{XXXXXXXX}\\
    \texttt{\_\_\_\_\_\_\_\_}\\
\end{tabular}&\begin{tabular}{c}
    \texttt{\_\_\_\_\_\_\_\_}\\
    \texttt{\_\_\_\_ZZ\_\_}\\
    \texttt{\_\_\_\_\_\_\_\_}\\
    \texttt{\_\_\_\_ZZ\_\_}\\
    \texttt{\_\_\_\_\_\_\_\_}\\
    \texttt{\_\_\_\_ZZ\_\_}\\
    \texttt{\_\_\_\_\_\_\_\_}\\
    \texttt{\_\_\_\_ZZ\_\_}\\
\end{tabular}&\begin{tabular}{c}
    \texttt{\_\_\_\_\_\_\_\_}\\
    \texttt{\_\_\_\_\_\_\_\_}\\
    \texttt{\_\_\_\_\_\_\_\_}\\
    \texttt{\_\_\_\_\_\_\_\_}\\
    \texttt{\_\_\_\_\_\_\_\_}\\
    \texttt{\_\_\_\_\_\_\_\_}\\
    \texttt{Z\_Z\_Z\_Z\_}\\
    \texttt{Z\_Z\_Z\_Z\_}\\
\end{tabular}&\texttt{{\kern 5em}}&\texttt{{\kern 5em}}&\begin{tabular}{c}
    \texttt{Z\_Z\_\_\_\_\_}\\
    \texttt{\_\_\_\_\_\_\_\_}\\
    \texttt{\_\_\_\_\_\_\_\_}\\
    \texttt{\_\_\_\_\_\_\_\_}\\
    \texttt{\_\_\_\_\_\_\_\_}\\
    \texttt{\_\_\_\_\_\_\_\_}\\
    \texttt{Z\_YX\_\_\_\_}\\
    \texttt{\_\_\_\_\_\_XX}\\
\end{tabular}&\texttt{{\kern 5em}}&\begin{tabular}{c}
    \texttt{Z\_\_\_Z\_\_\_}\\
    \texttt{\_\_\_\_\_\_\_\_}\\
    \texttt{\_\_\_\_\_\_\_\_}\\
    \texttt{\_\_\_\_\_\_\_\_}\\
    \texttt{\_\_\_\_\_\_\_\_}\\
    \texttt{\_\_\_\_\_\_\_\_}\\
    \texttt{Z\_\_\_YX\_\_}\\
    \texttt{\_\_\_\_\_\_XX}\\
\end{tabular}&\texttt{{\kern 5em}}&\begin{tabular}{c}
    \texttt{Z\_\_\_\_\_Z\_}\\
    \texttt{\_\_\_\_\_\_\_\_}\\
    \texttt{\_\_\_\_\_\_\_\_}\\
    \texttt{\_\_\_\_\_\_\_\_}\\
    \texttt{\_\_\_\_\_\_\_\_}\\
    \texttt{\_\_\_\_\_\_\_\_}\\
    \texttt{Z\_\_\_\_\_YX}\\
    \texttt{\_\_\_\_\_\_XX}\\
\end{tabular}&\texttt{{\kern 5em}}\\\hline
\begin{tabular}{c}
    \texttt{\_\_\_\_\_\_\_X}\\
    \texttt{\_\_\_\_\_\_\_X}\\
    \texttt{\_\_\_\_\_\_\_X}\\
    \texttt{\_\_\_\_\_\_\_X}\\
    \texttt{\_\_\_\_\_\_\_X}\\
    \texttt{\_\_\_\_\_\_\_X}\\
    \texttt{\_\_\_\_\_\_\_X}\\
    \texttt{\_\_\_\_\_\_\_X}\\
\end{tabular}&\begin{tabular}{c}
    \texttt{\_\_\_\_\_\_\_\_}\\
    \texttt{\_\_\_\_\_\_\_\_}\\
    \texttt{\_\_\_\_\_\_\_\_}\\
    \texttt{\_\_\_\_\_\_\_\_}\\
    \texttt{\_\_\_\_\_\_\_\_}\\
    \texttt{\_\_\_\_\_\_\_\_}\\
    \texttt{\_\_\_\_\_\_\_\_}\\
    \texttt{XXXXXXXX}\\
\end{tabular}&\begin{tabular}{c}
    \texttt{\_\_\_\_\_\_\_\_}\\
    \texttt{\_\_\_\_\_\_ZZ}\\
    \texttt{\_\_\_\_\_\_\_\_}\\
    \texttt{\_\_\_\_\_\_ZZ}\\
    \texttt{\_\_\_\_\_\_\_\_}\\
    \texttt{\_\_\_\_\_\_ZZ}\\
    \texttt{\_\_\_\_\_\_\_\_}\\
    \texttt{\_\_\_\_\_\_ZZ}\\
\end{tabular}&\begin{tabular}{c}
    \texttt{\_\_\_\_\_\_\_\_}\\
    \texttt{\_\_\_\_\_\_\_\_}\\
    \texttt{\_\_\_\_\_\_\_\_}\\
    \texttt{\_\_\_\_\_\_\_\_}\\
    \texttt{\_\_\_\_\_\_\_\_}\\
    \texttt{\_\_\_\_\_\_\_\_}\\
    \texttt{\_Z\_Z\_Z\_Z}\\
    \texttt{\_Z\_Z\_Z\_Z}\\
\end{tabular}&\texttt{{\kern 5em}}&\texttt{{\kern 5em}}&\begin{tabular}{c}
    \texttt{Z\_Z\_\_\_\_\_}\\
    \texttt{\_\_\_\_\_\_\_\_}\\
    \texttt{\_\_\_\_\_\_\_\_}\\
    \texttt{\_\_\_\_\_\_\_\_}\\
    \texttt{\_\_\_\_\_\_\_\_}\\
    \texttt{\_\_\_\_\_\_\_\_}\\
    \texttt{\_\_\_\_\_\_\_\_}\\
    \texttt{Z\_YX\_\_XX}\\
\end{tabular}&\texttt{{\kern 5em}}&\begin{tabular}{c}
    \texttt{Z\_\_\_Z\_\_\_}\\
    \texttt{\_\_\_\_\_\_\_\_}\\
    \texttt{\_\_\_\_\_\_\_\_}\\
    \texttt{\_\_\_\_\_\_\_\_}\\
    \texttt{\_\_\_\_\_\_\_\_}\\
    \texttt{\_\_\_\_\_\_\_\_}\\
    \texttt{\_\_\_\_\_\_\_\_}\\
    \texttt{Z\_\_\_YXXX}\\
\end{tabular}&\texttt{{\kern 5em}}&\texttt{{\kern 5em}}&\texttt{{\kern 5em}}\\\hline
\end{tabular}

    }
    \caption{
        Definitions of stabilizers and observables in the $n=64,k=34,d=4$ 2D parity check code.
        Each cell shows the entries of a stabilizer or observable as an $8\times8$ grid of Pauli terms; one term for each of the 64 physical qubits.
        Note that both the X observables and Z observables can be recovered from the Y observables, as the X observables contain no Y or Z terms and the Z observables contain no X or Y terms.
        Also note that two of the listed checks are redundant - the product of all X col checks is equal to the product of all X row checks, and the product of all Z bi-col checks is equal to the product of all Z bi-row checks.
    }
    \label{tab:squareberg_code}
\end{table}

More generally, we require an $r$-dimensional parity check code to have each side length divisible by $2^r$, see \app{mdpc}.
An $r$-dimensional parity check code has distance $2^r$, with minimum weight logical operators forming the vertices of $r$-dimensional rectangles.
We can count the number of independent checks to determine it has parameters $[[\prod n_i, 2\prod(n_i - 1) - \prod n_i, 2^r]]$, where $n_i$ are the side lengths of the $r$-dimensional array. 
In particular, for 1D and square 2D codes, we obtain families of $[[n, n-2, 2]]$ \cite{steane1996simpleqec} and $[[n^2, n^2 - 4n + 2, 4]]$ codes, respectively.

\section{Lattice surgery constructions}
\label{sec:lattice_surgery}

A key detail when concatenating a code over the surface code is how the checks of the overlying code will be measured.
This is important as the workspace needed to periodically measure the checks of the overlying code can easily be larger than the space needed to just store the qubits.
Furthermore, we must account for the effects of errors occurring during the lattice surgery in order to ensure fault-tolerance of the outer code.

Originally, lattice surgery was thought of as being built out of parity measurements~\cite{horsman2012latticesurgery,fowler2018latticesurgery}, but a far more useful perspective is to view lattice surgery as an instantiation of the ZX calculus~\cite{de2017zxlattice}.
In the latter perspective, the building blocks of lattice surgery are not operations on qubits but rather connections between junctions.
Making a good lattice surgery construction then becomes an exercise in packing, routing, and rotating pipes so that they link together in the required way.

\begin{figure}[h]
    \centering
    \resizebox{\linewidth}{!}{
     \includegraphics{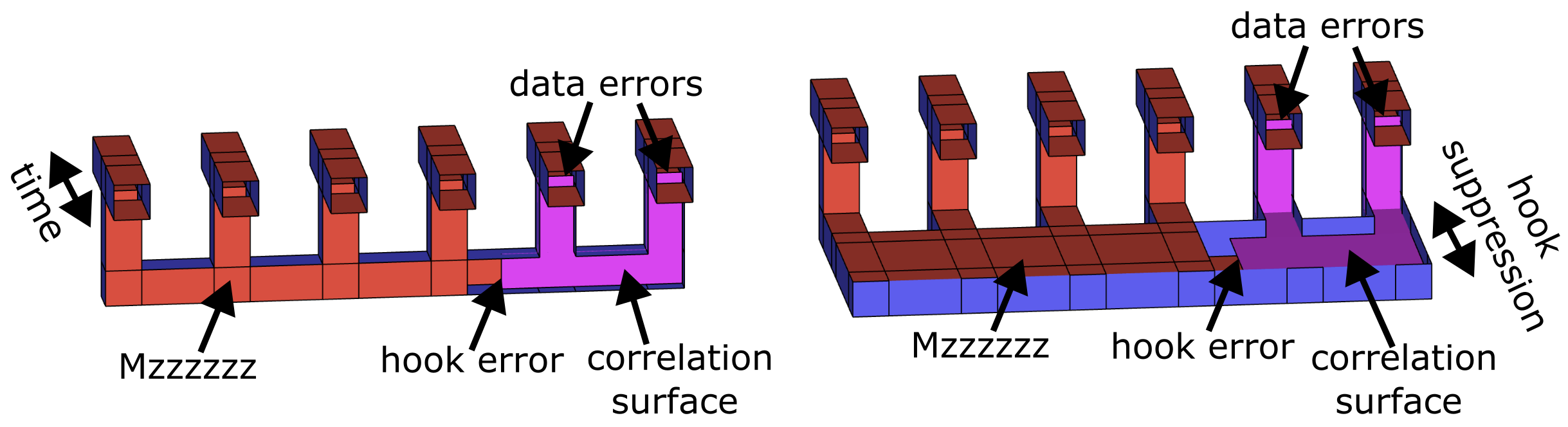}
    }
    \caption{
        Topological diagrams of multi-body logical measurements, with connections between blocks stretched out to show the topology.
        Left: a multi-body $Z$-measurement. The correlation surface shows the equivalence of a short spacelike ``hook'' error to two data errors.
        Right: the same multi-body $Z$-measurement with protection against correlated hook errors.
        We can increase protection against the hook error by extending the distance between the boundaries that it connects.
        Naively, this would increase the overall qubit footprint of the circuit.
        However, we can orient this extension in time, trading a smaller qubit footprint for a longer syndrome extraction cycle.
    }
    \label{fig:long_stabilizers}
\end{figure}

\begin{figure}[h]
    \centering
    \resizebox{\linewidth}{!}{
        \includegraphics{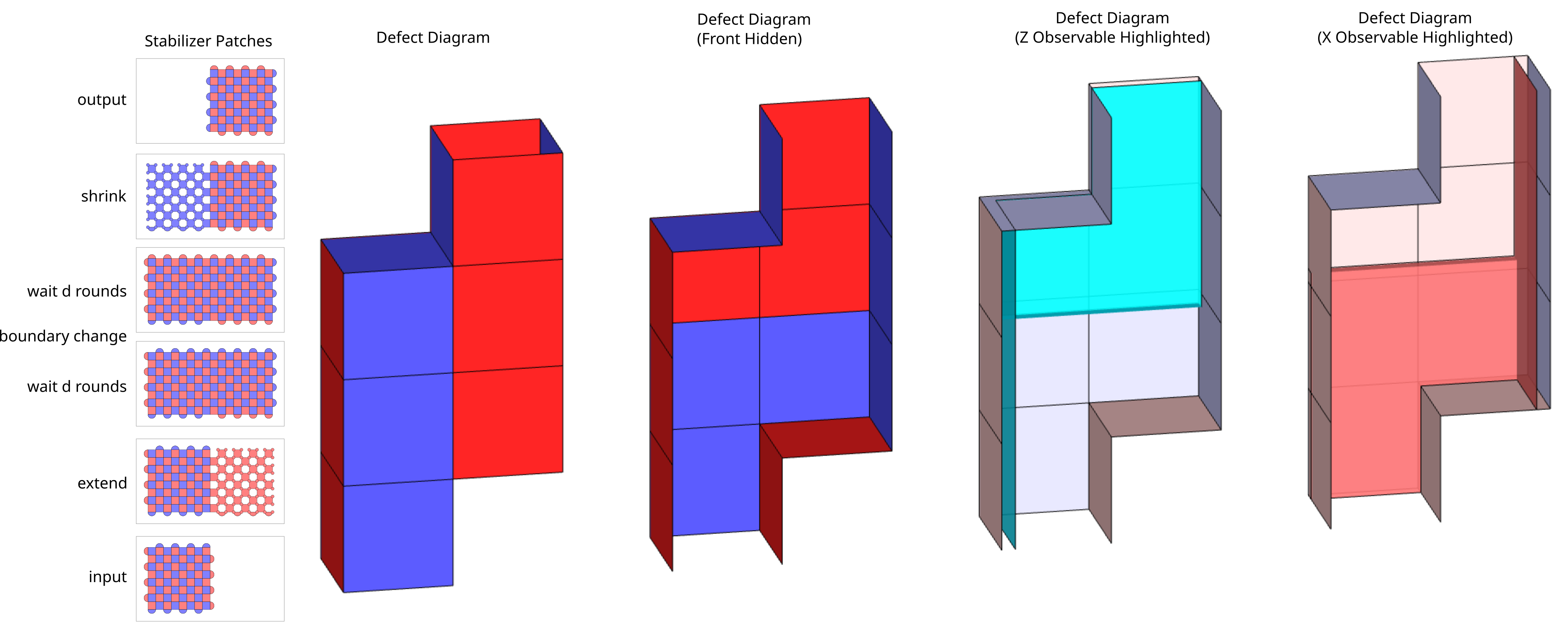}
    }
    \caption{
        The patch rotation construction from \cite{litinski2019gameofsurfacecodes}, with the last step omitted, leaving the patch shifted as part of the rotation.
    }
    \label{fig:patch_rotation}
\end{figure}

In \fig{long_stabilizers}, we illustrate example multi-qubit measurements performed using lattice surgery, and in \fig{patch_rotation}, patch rotations that we will use to glue different measurements together.
Typically, when executing a fault-tolerant circuit described in terms of these topological diagrams, we don't concern ourselves with which error strings occur.
We assume that any error string will corrupt the circuit.
However, when concatenating the surface code into an outer code, we must worry about error propagation in much the same way we do when designing fault-tolerant syndrome extraction at the base code level.
One disadvantage of using high-density parity check codes is that measuring these high-weight stabilizers can induce correlated failures among the inner surface codes, which might not be corrected by the outer code.
These are analogous to hook errors propagating from a measure qubit to many data qubits.

However, unlike physical qubits, surface codes can modulate the distances between boundaries to bias protection against different error mechanisms.
Of course, this might enlarge the footprint of the concatenated code by extending the size of the base code.
However, a second nice property of the surface code is that its topological operations are mostly agnostic to their spacetime orientation.
Consequently, we can orient this extended protection in the time direction, holding the spatial footprint of the yoked surface codes fixed.
For example, in \fig{long_stabilizers}, we orient the correlated error in time and extend the protection against it to maintain the concatenated codes effective doubled distance.
This isn't a free lunch: it increases the length of the outer code's syndrome cycle, which in turn increases the distance required by the inner code.
However, as the error rate scales polynomially with the length of the syndrome cycle and inverse exponentially with the distance of the inner code, this tradeoff is not too damaging.

As 1D and 2D yoked surface codes have doubled and quadrupled code distances respectively, we elect to extend the duration of the yoke checks to ensure each correlated error is protected to distance $2d_{\text{inner}}$ (as in \fig{long_stabilizers}) and $4d_{\text{inner}}$, respectively.
This is likely overly conservatives, as it effectively suppresses the 1D and 2D correlated failures using $2d_{\text{inner}}$ and $4d_{\text{inner}}$ (asymmetric) surface codes, which we will see provide significantly better per-distance protection than yoked surface codes on uncorrelated failures.
This makes the probability of a correlated failure negligible relative to the dominant uncorrelated failure mechanism, and we could likely shorten the syndrome cycle considerably.
However, we err on the side of caution in our estimates, and use overly protected syndrome measurements.
Note that this does provide more opportunities for spacelike failure paths, which corresponds to increasing the effective measurement error probability of the outer code.

\begin{figure}[h]
    \centering
    \resizebox{\linewidth}{!}{
        \includegraphics{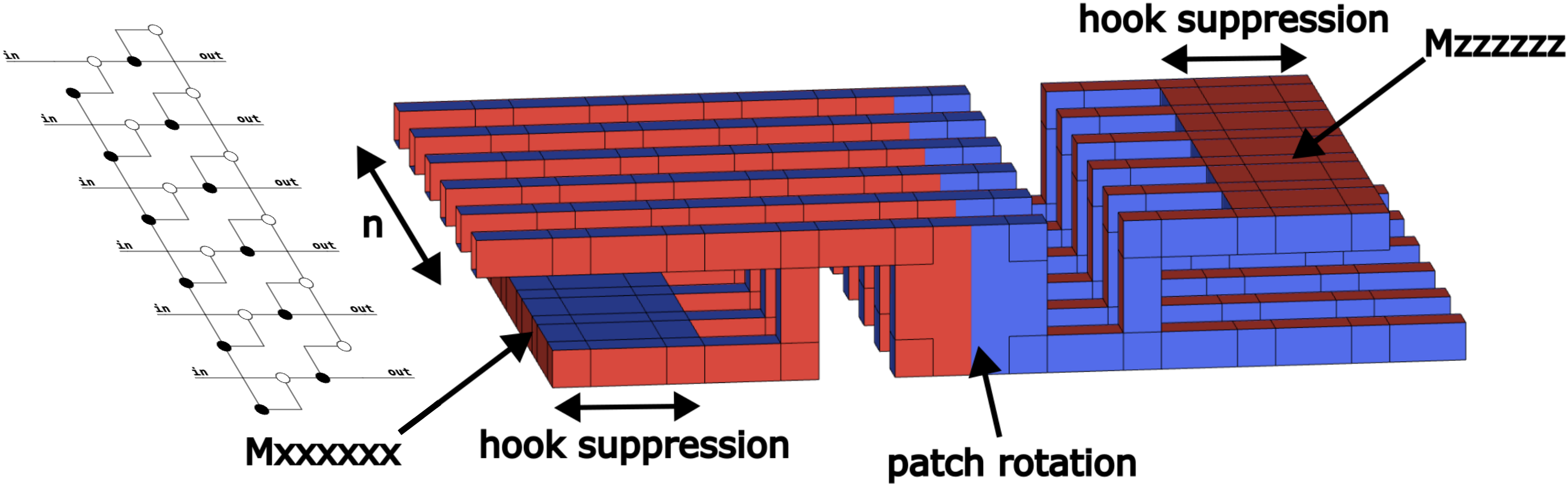}
        }
    \caption{
        Checking the $X^{\otimes n}$ and $Z^{\otimes n}$ stabilizers of 1D yoked surface codes using lattice surgery.
        The process occupies $2n$ surface code patches for $8d$ rounds, where $n$ is the block length of the outer code. 
        Time flows left to right.
        The corresponding ZX diagram is shown to the left.
    }
    \label{fig:1d_yoke_pipe_component}
\end{figure}

\begin{figure}[h]
    \centering
    \resizebox{\linewidth}{!}{
        \includegraphics{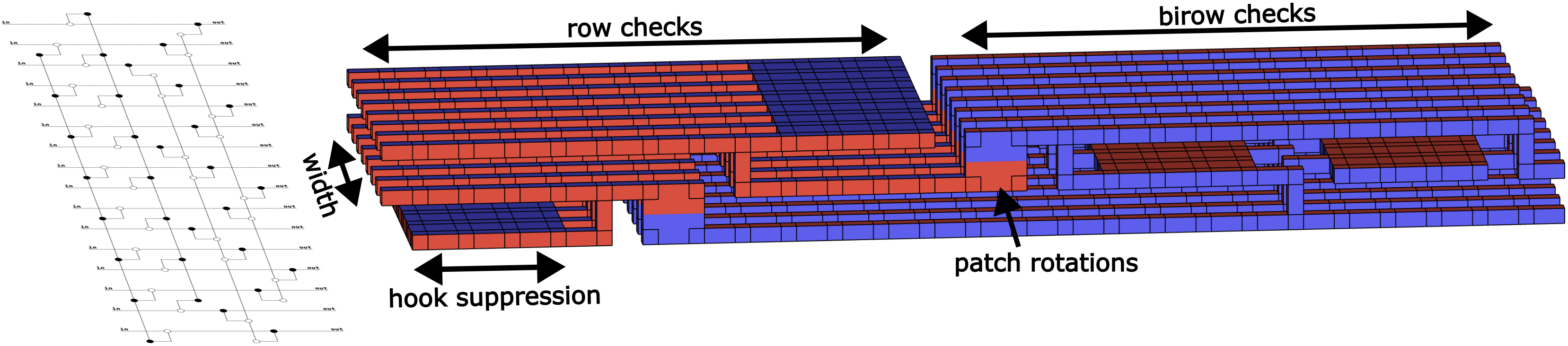}
        }
    \caption{
        Checking two row and birow stabilizers of 2D yoked surface codes using lattice surgery.
        This process occupies $3w$ surface code patches for $25d$ rounds, where $w = \sqrt{n}$ is the the width of the outer array. 
        Time flows left to right. 
        The corresponding ZX diagram is shown to the left.
        In that diagram, top pipes correspond to every other wire beginning from the top, while bottom pipes correspond to every other wire beginning second from the top.
    }
    \label{fig:2d_yoke_pipe_component}
\end{figure}

\begin{figure}[h]
    \centering
    \resizebox{\linewidth}{!}{      \includegraphics{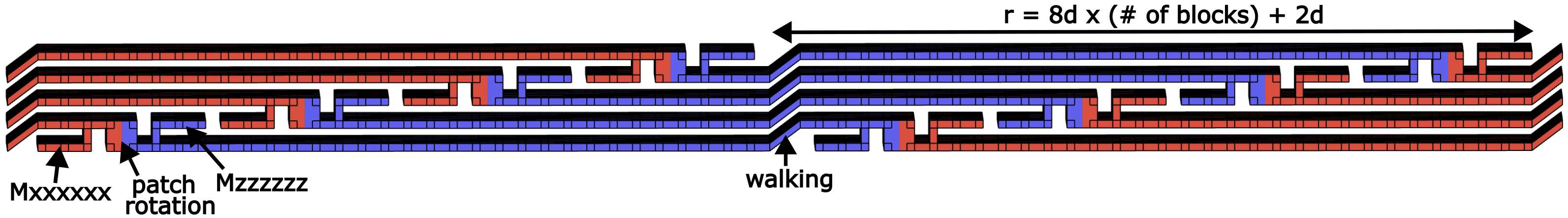}
    }
    \caption{
        The full syndrome extraction of four blocks of 1D yoked surface codes.
        An extra row of workspace surface code patches travels through the blocks to measure the outer stabilizers. We use the walking surface code construction in \cite{mcewen2022relaxing} to connect the outer syndrome cycles, which takes $2d$ rounds to execute.
        The total length of a syndrome cycle scales as $8d \times \text{(\# of blocks)} + 2d$.
    }
    \label{fig:1d_full_cycle}
\end{figure}

\begin{figure}[h]
    \centering
    \resizebox{\linewidth}{!}{      
    \includegraphics{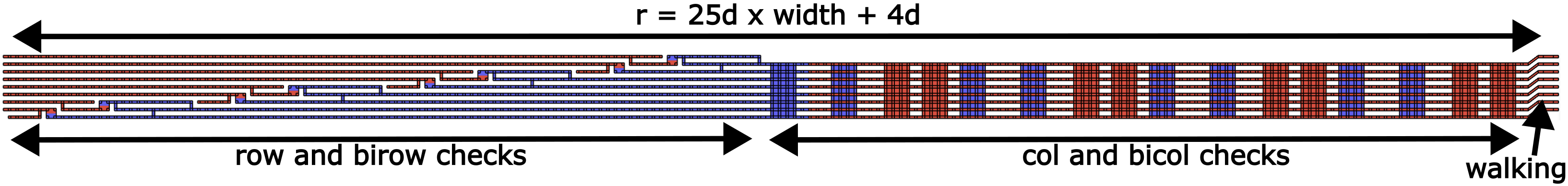}
    }
    \caption{
        The full syndrome extraction of 2D yoked surface codes, with an outer $[[64,34,4]]$ code.
        An extra row and column of workspace surface code patches travels through the blocks to measure the outer stabilizers. We use two iterations of walking surface codes to connect the outer syndrome cycles, which together take $4d$ rounds to execute.
        The total length of a syndrome cycle scales as $25dw + 4d$, where $w = \sqrt{n}$ is the width of the square outer code array, in this case $8$.
    }
    \label{fig:2d_full_cycle}
\end{figure}

To construct the full syndrome extraction circuit, we build it up from pieces.
In \fig{1d_yoke_pipe_component} and \fig{2d_yoke_pipe_component}, we show the $X$- and $Z$-type outer stabilizers measured by combining parity measurements with patch rotations.
Finally, we fit these puzzle pieces together to form the full syndrome cycle circuits in \fig{1d_full_cycle} and \fig{2d_full_cycle}.
Note that there is extra workspace required to measure these checks.
For example, in 1D yoked surface codes, we use a single workspace row to sequentially measure several 1D yoked surface code blocks, analogous to a measure qubit migrating across the code blocks to extract stabilizer measurements.
Having a single extra row attend to all the blocks lengthens the outer code's syndrome extraction cycle, but reduces the number of occupied surface code patches.
Striking a balance between these effects is important, as we must include the overhead of this workspace in the yoked surface code's overall footprint.

We also consider a second storage format, which we call ``hot storage''.
In our previous format ``cold storage'', logical qubits were stored as densely as possible, but could not be immediately operated upon.
Operating on a logical qubit in cold storage requires first getting it out of storage.
Concretely, this means the surface code patches don't all have access hallways available next to them.
It's still necessary for some workspace to be present, because it's necessary to periodically check the yokes, but this workspace may be shared between many groups of patches.

\begin{figure}[htb!]
    \centering
    \resizebox{\linewidth}{!}{
        \includegraphics{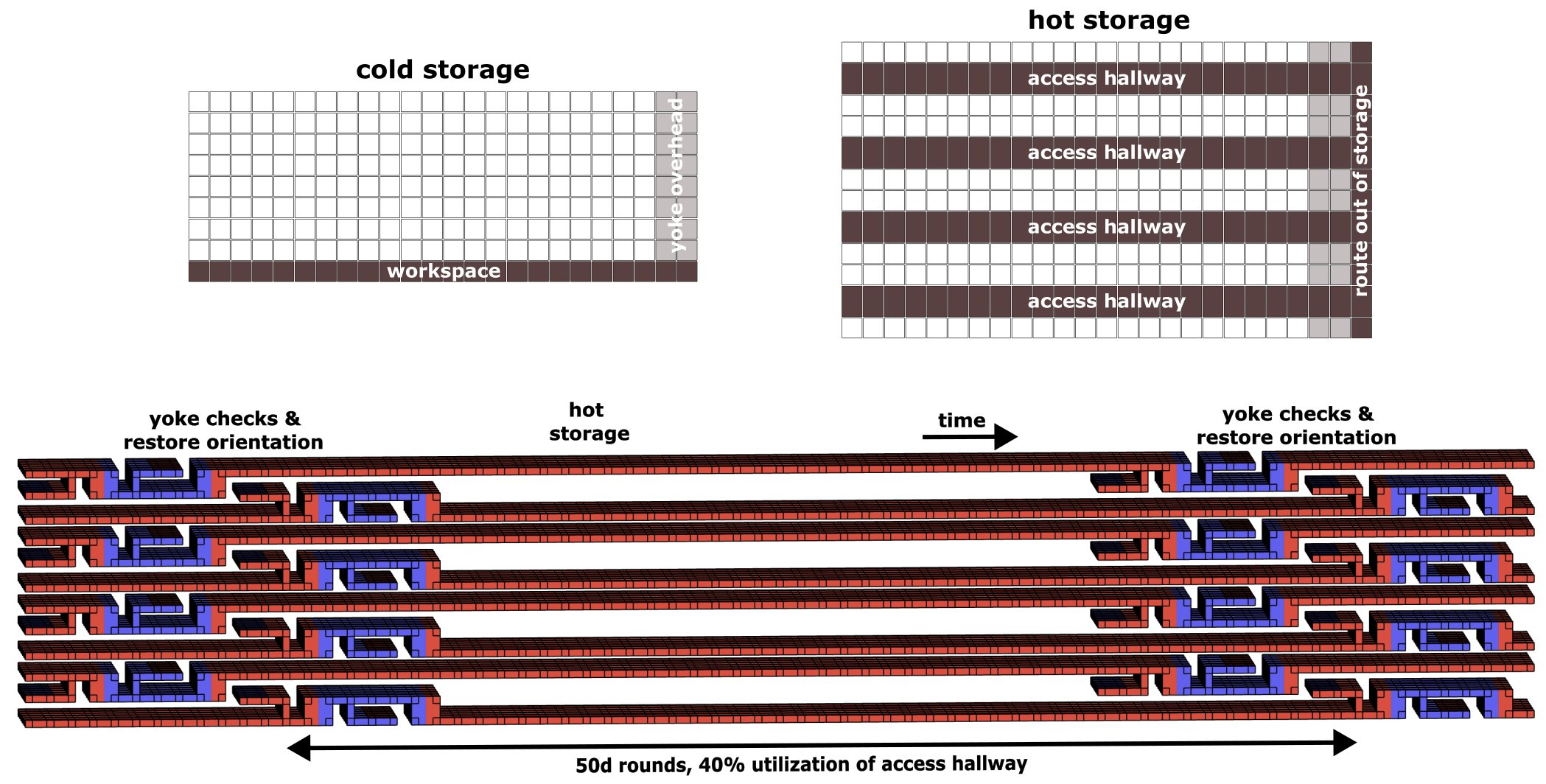}
    }
    \caption{
        Layouts for cold and hot storage using 1D yoked surface codes.
        The 2D footprint diagrams at the top show how space is allocated, while the 3D topological diagrams at the bottom show hot storage syndrome cycles occurring over time.
        In the 2D footprint diagrams, each row is a separate outer code block.
        The white-filled squares correspond to usable storage while other squares correspond to various overheads.
        In cold storage, one row of workspace is shared between different blocks in order to measure the yokes.
        In hot storage, the yokes are measured using the access hallways that are already present.
        We assume that the access hallways are the full unyoked code distance in height.
        The hot storage syndrome cycle shown takes 50$d$ rounds and utilizes the access hallways 40\% of the time.
    }
    \label{fig:storage}
\end{figure}

Logical qubits in hot storage are available to be operated upon.
Each surface code patch has a boundary exposed to an access hallway that provides a route out of storage.
The existence of this access hallway is useful for yoked surface codes because it can also be used as a workspace for periodically measuring the yokes.
This means measuring the yokes has no marginal space cost; the space was already paid for.
Instead, it has a marginal spacetime cost, because although the access hallway was already there, it's blocked while a yoke is being measured.
\fig{storage} shows the space layout, and spacetime layout, that we use for estimating the overhead of hot storage.

In this work, we only consider hot storage of 1D yoked surface codes.
We expect the access hallway requirements and access hallway utilization of 2D yoked surface codes to be too demanding.
Storage could be even hotter than we consider here, by having each surface code patch expose two boundaries~\cite{litinski2019gameofsurfacecodes,litinskyhypercubeshor2023}.
We consider the additional space cost of exposing two boundaries as less worth the benefit, and so don't consider layouts of that type.
Note that this means we're assuming that qubits in hot storage are rotated on an as-needed basis, when we need to access the boundary that isn't exposed.
The need to perform these rotations is a key consideration when laying out an algorithm.

Yoked surface codes in hot storage can be operated on by lattice surgery while they're encoded.
Each encoded observable is spread over several surface code patches, but lattice surgery can stitch to several patches as easily as one.
The cost is actually essentially identical to doing lattice surgery with unyoked surface codes, because the entrance to the access hallway is occupied regardless of how many patches are touched.
However, we must be careful when operating on these encoded qubits and ensure that we do not frequently expose unprotected lower-distance patches.
Consequently, to keep our overhead estimates conservative, we assume access hallways require the full unyoked code distance in height.

Evidently, using yoked surface codes in fault-tolerant computation has consequences on the large scale architecture of a quantum computer, beyond just the size of the storage.
Operating on yoked surface codes is important to understand, but beyond the scope of this paper.

\section{Complementary gaps}
\label{sec:complementary_gaps}

From the perspective of the outer code, the syndrome of the inner code gives valuable information about the likelihood of an error in a particular location.
For example, an inner surface code with no detection events is far less likely to have experienced an error than an inner surface code with many detection events.
We can quantify this likelihood by comparing the probability of a set of errors obtained from minimum-weight matching against the probability of a set of errors obtained from a minimum-weight matching conditioned on the complementary logical outcome.
Passing this information to the outer code helps it to identify likely culprit errors \cite{poulin2006optimal}.

Operationally, we can compute the minimum-weight matching conditioned on the complementary logical outcome by modifying the error graph.
Given a block of surface code memory with boundaries, we can form a detector connecting to all the boundary edges on one side of the error graph, similar to \cite{hutter2014complementarygap}.
This augmentation maintains the graph structure, and turning this boundary detector on/off forces the decoder to match/not match to the corresponding boundary.\footnote{We note that an upcoming work \cite{meister2023efficient} presents a different highly efficient and flexible soft decoding method.}
The resulting two matchings are the decoder's best hypotheses for the set of errors explaining these two topologically distinct classes of errors.
We call the log-likelihood ratio of these two hypotheses the complementary gap - the log-ratio of the probabilities of the minimum-weight matching and the complementary matching.
A complementary gap close to zero indicates that the decoder is not confident in its decision, while a high complementary gap indicates the decoder is highly confident.

This information is extremely helpful and can be used in decoding the outer code.
1D and 2D parity check codes can themselves be decoded using minimum-weight perfect matching, and so we use these complementary gaps as edge weights in the outer error graph.
These edge weights represent the cost of flipping a minimum-weight matching to a complementary matching.
There are several ways to generalize this procedure to a correlated matching decoder.
In this work, we use a two-pass correlated matching decoder similar to the one described in \cite{fowler2013optimal}.
To compute the complementary gap in a $Z$-basis memory experiment, we use the $X$-type error graph to reweight the $Z$-type error graph, and then compute the complementary gap for this reweighted $Z$-type error graph.

\begin{figure}[h]
    \centering
    \resizebox{0.75\linewidth}{!}{    
    \includegraphics{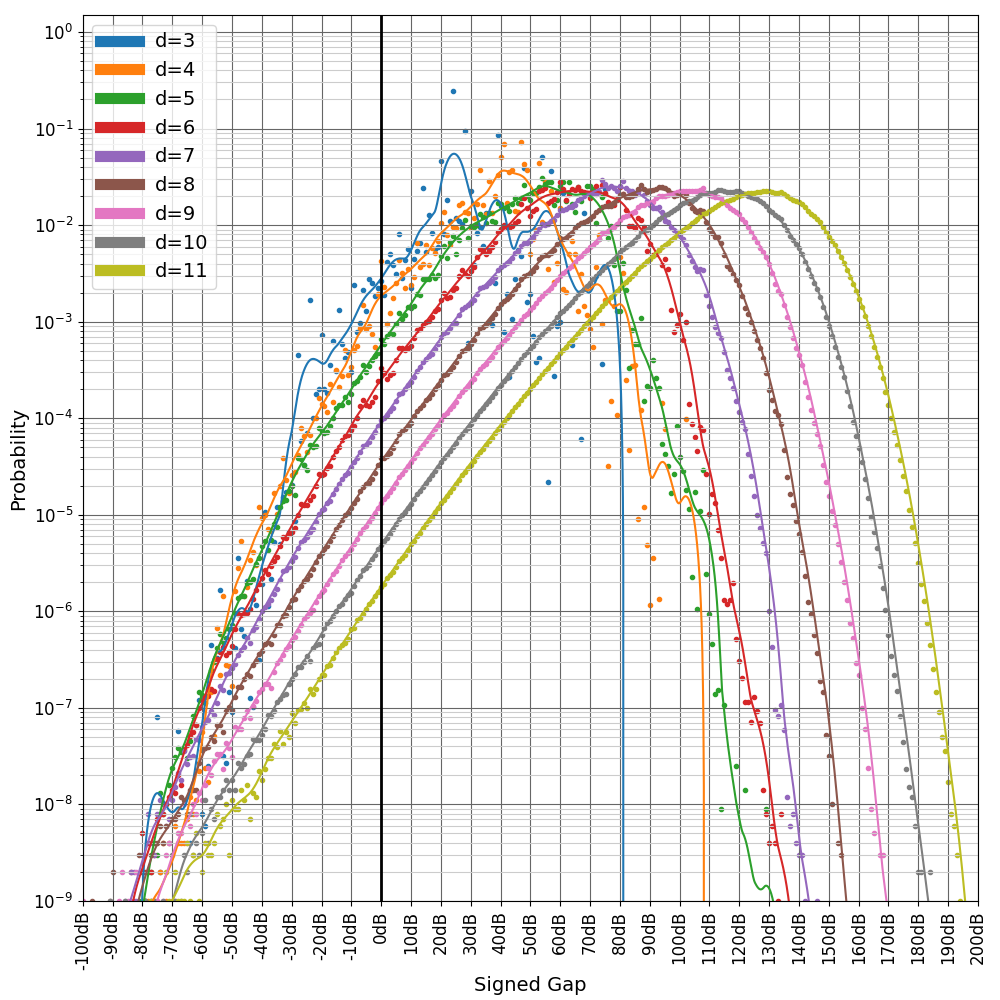}
    }
    \caption{
        Distributions of sampled complementary gaps for $10d$ round memory experiments with perfect terminal time boundaries checking one observable.
        This data was gathered using the SI1000 error model described in \app{noise} at an error rate of $10^{-3}$.
        Gaps are presented in terms of their ratio in dB, where each gap is binned into the nearest integer dB.
        A negative gap indicates that the more likely outcome was incorrect.
        Probability distributions are smoothed by convolving with a cosine window and rescaled so that the smoothed output has the same area under the curve as the input.
        Overall, each curve is comprised of $10^9$ samples.
    }
    \label{fig:raw_gap}
\end{figure}

\begin{figure}[h]
    \centering
    \resizebox{\linewidth}{!}{    
    \includegraphics{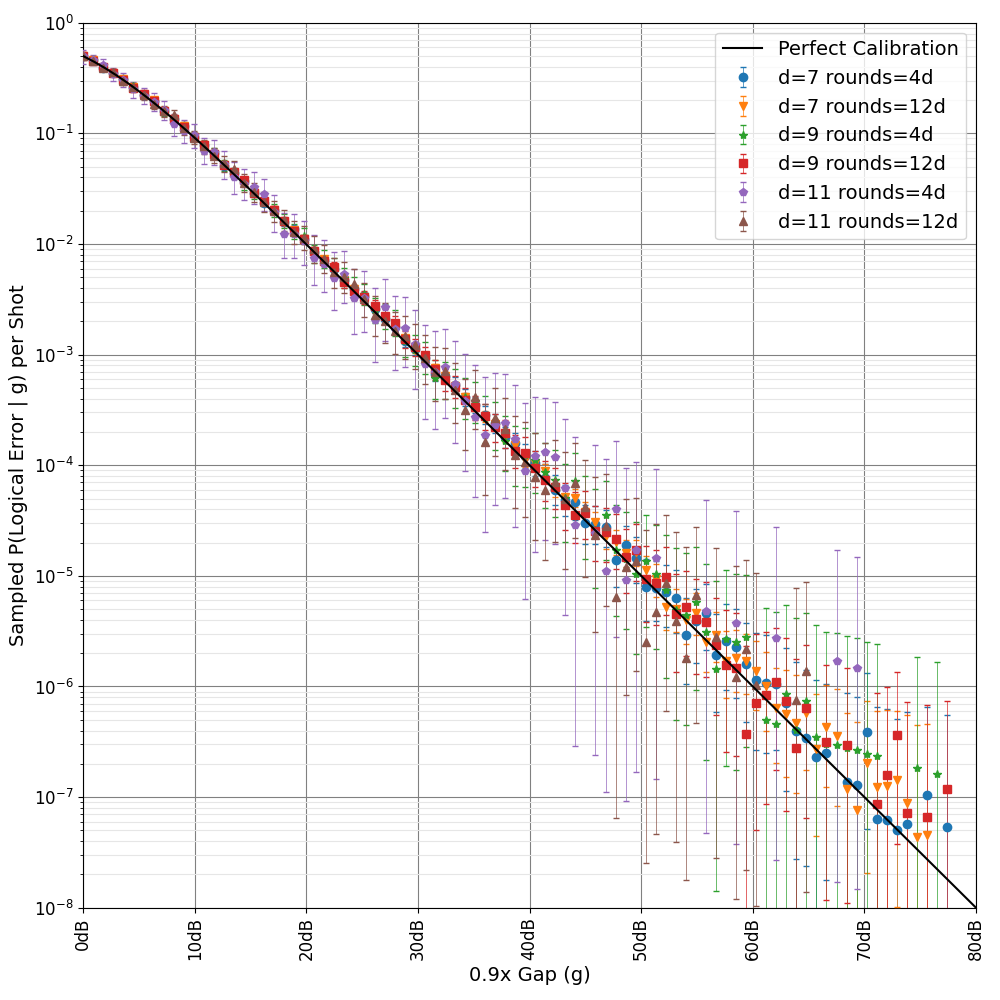}
    \includegraphics{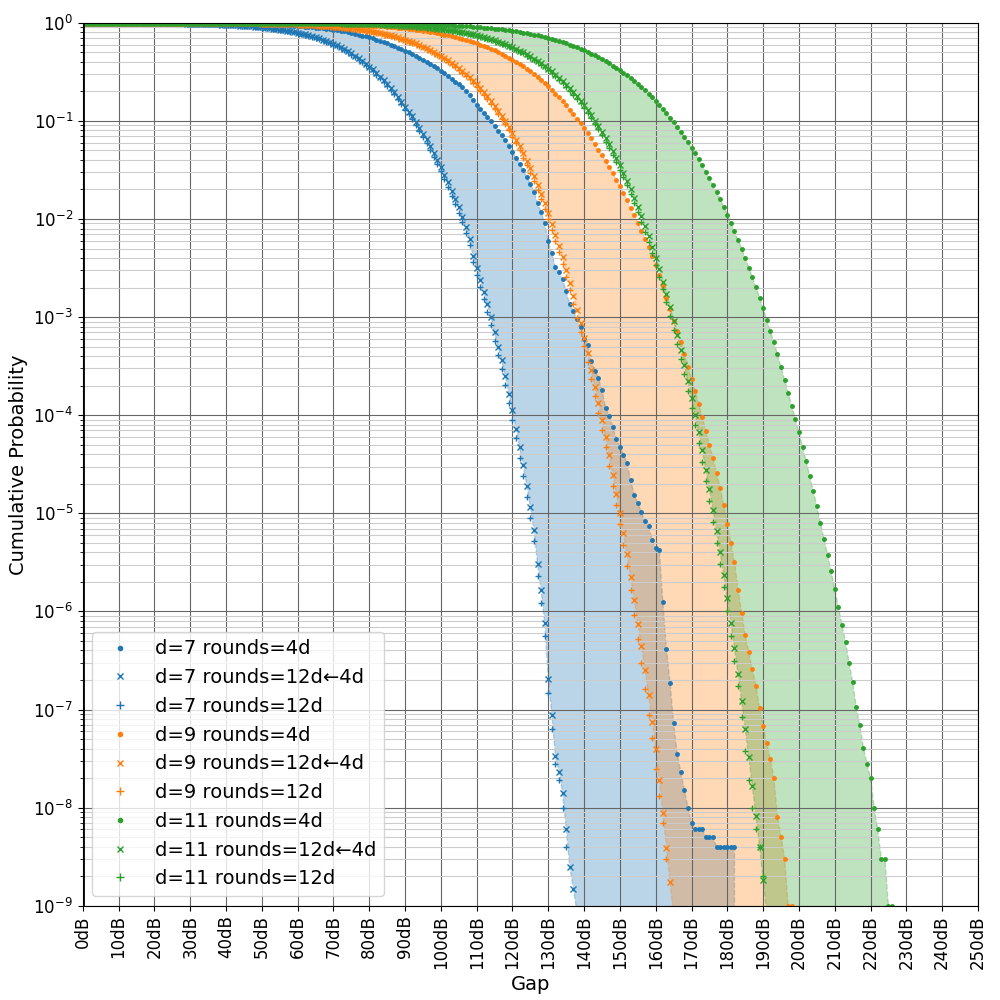}
    }
    \caption{
        Left: the decoder calibration after rescaling by 0.9x. 
        The decoder is well-calibrated across different distances and rounds, but tends towards overconfidence in its prediction.  
        Right: extrapolating the inverse cumulative distribution functions of the complementary gaps by exponentiating.
        This data was gathered using the SI1000 error model described in \app{noise} at an error rate of $10^{-3}$.
        Extrapolations are represented as x's with the intervening space shaded.
        We observe that this well-approximates the inverse cumulative distribution function of a longer memory experiment, with a slight tendency towards sampling too-large gaps as the gap increases.
    }
    \label{fig:gap}
\end{figure}

In \fig{raw_gap}, we see that the gap distributions take a smooth, simple form after an initially noisy start at low distance, likely due to finite-size effects.
For our benchmarks, it will be important to extrapolate the behavior of these gaps, as we will use them to estimate the behavior of very large simulations at low error rates - see \fig{gap}.
We observe that the gaps are well-calibrated - the likelihood of success predicted by the gap is close to the true empirical likelihood of success after rescaling the gap by 0.9x.
That is, we rescale the decoder's confidence to account for its slight over-confidence in high-confidence predictions.
The decoder also remains well-calibrated when extended over many rounds.
We also observe the distribution on complementary gaps over $mn$ rounds can be well-approximated as the minimum of $m$ samples from the distribution on gaps over $n$ rounds.
Although these approximations tend to be slightly optimistic, they allow us to extrapolate the probability of observing a particular gap and the resulting likelihood of failure from a distribution of gaps on relatively few (e.g. a small multiple of $d$) rounds.

\section{Benchmarking}
\label{sec:benchmarking}

Simulating yoked surface codes introduces some difficulties.
We want to probe very low error rates on blocks of very many logical qubits.
Consequently, we perform two types of simulations.

The first type of simulation is a circuit-level simulation of the inner code over a relatively small number of inner rounds, with a single perfect outer round of yoke checks.
Yoke detectors are placed along the boundaries of the inner surface code error graphs, with yoke detectors occupying either one or two boundaries of the error graph in the case of 1D or 2D outer codes.
This turns decoding of the concatenated code into a single minimum-weight matching problem.
Note that, when casting concatenated decoding as a single matching problem, the yoke detectors must connect to the boundary.
Otherwise, individual errors might introduce more than two detection events on a particular error graph.

For longer simulations, we phenomenologically simulate multiple rounds of the outer codes using the gap distributions described in \sec{complementary_gaps}.
The effective error rate seen by the outer code is determined by the distance of the inner code, as well as the approximate spacetime volume that contributes to a particular error edge.
We consider spacelike and timelike edges only.
While some of the spacelike error probability may be redistributed to spacetime-like edges in 2D outer codes (depending on different lattice surgery schedules), we do not expect this added complexity to introduce a significant effect.
In fact, we will see that timelike edges do not contribute significantly to the overall error rate, which appears to be dominated by shortest-path error configurations.
We also ignore damaging hook errors, having already paid to suppress them below the relevant noise floor by using the hook suppression constructions in \sec{lattice_surgery}.

These simulations proceed as follows.
First, we simulate the complementary gap distribution over $10^9$ shots and varying code distances at $10d$ rounds of a $Z$-type memory experiment - these distributions (binned by nearest integer dB) are recorded in \fig{raw_gap}.
We record the probability of observing a particular gap as well as the likelihood that particular gap results in an error, i.e. that the minimum-weight matching does not belong to the same error class as the true error configuration.
Frequently, we might sample gaps for which we have seen no failures.
Extrapolating from the calibration curve in \fig{gap}, we assign these events a likelihood of failure corresponding to 0.9x the sampled gap.
For each edge in the outer error graph, we associate a number of rounds $N$ that contribute to an error resulting in flipping that edge.
Heuristically, we focus on the time extent between rounds to determine $N$.
Some components (like patch rotations) can have increased spatial extent, but also frequently come with relatively fewer shortest paths or average out against increased protection during a later time step.
We treat timelike outer edges separately, counting rounds according to their spacetime extent, since they extend much farther in space than time. 
To extrapolate the complementary gap distributions, we raise the inverse cumulative distribution function to the $\frac{N}{10d}$ power, corresponding to taking the minimum gap over $\frac{N}{10d}$ draws from the $10d$ round complementary gap distribution.

Finally, we perform our simulations by sampling, for each edge in the isomorphic $Z$- and $X$-type error graphs, the complementary gap and the corresponding likelihood of failure.
In 1D yoked surface codes, these are timelike line graphs with multiple boundary edges at each node.
In 2D yoked surface codes, these are layers of complete bipartite graphs between row check detectors and column check detectors, with timelike edges connecting the layers.
We sample gaps to determine which edges have suffered an error, and use those gaps to weight the edges.
Finally, we decode these weighted outer error graphs using minimum-weight matching.
Note that separately simulating and decoding the $Z$- and $X$-type error graphs could negate useful correlations between them, but these benefits are likely small given the surface code's bias against logical $Y$-type errors (see \fig{y-bias}).

\begin{figure}[htb!]
    \centering
    \resizebox{\linewidth}{!}{
     \includegraphics{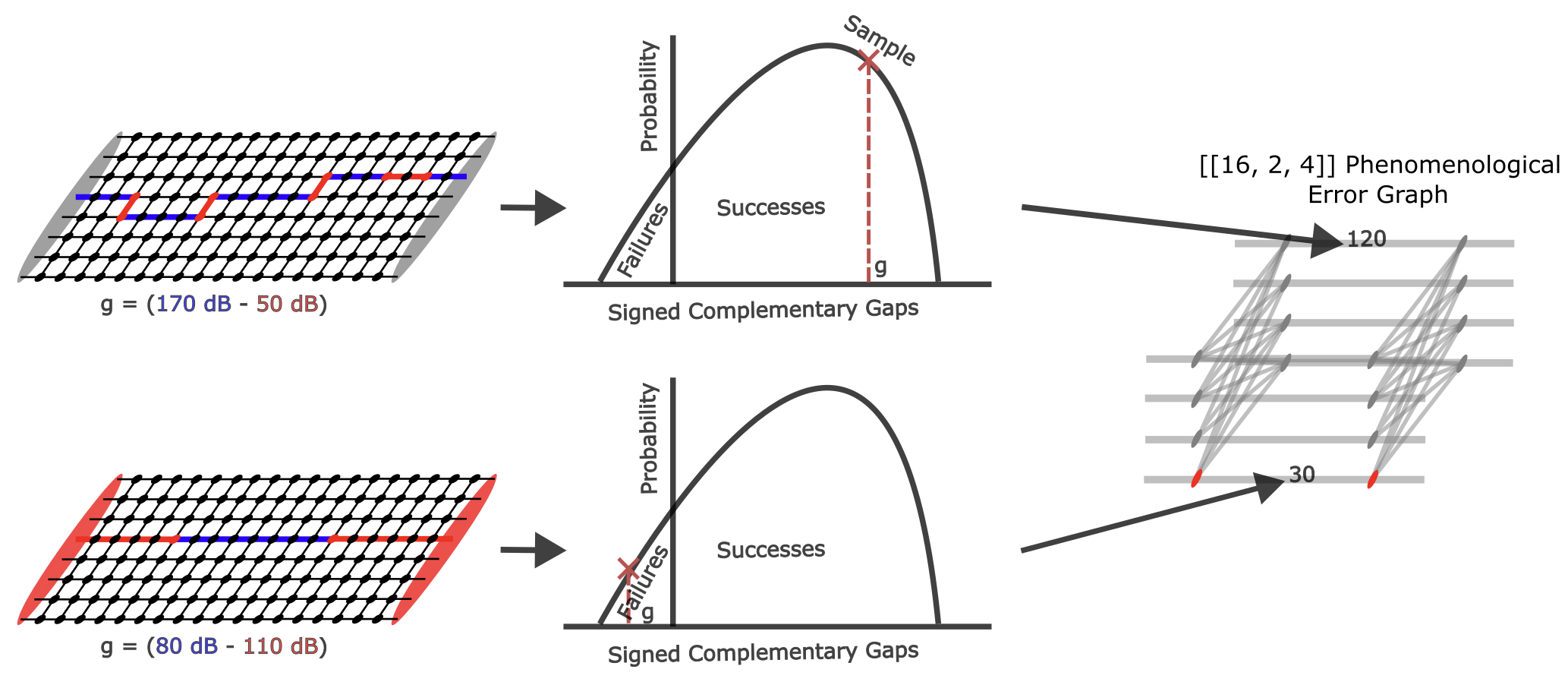}
    }
    \caption{
        Simplified example of decoding an outer code abstracted from complementary gaps.
        In the example, each edge of the left-side inner error graph has a relative probability of $-10dB$ of erring.
        Red edges denote the true errors, which are predicted by the minimum-weight matching on the top but not the bottom.
        Boundary ovals represent yoke detectors, some of which are activated.
        We replace a full circuit-level simulation with sampling signed gaps.
        The gap distributions are modified to extrapolate to the number of rounds an outer edge represents.
        Negative gaps indicate the minimum-weight matching wrongly identified the error class.
        These signed gaps are then used to populate the outer edge weights and detection events of the outer error graph, which is then also decoded using minimum-weight matching.
    }
    \label{fig:complementary_gap}
\end{figure}

\subsection{Scaling approximations}
In order to flexibly choose the inner code distance, code block size, and number of code blocks to meet a target logical error rate, our simulations validate simple, heuristic scaling laws that we can use to estimate the logical error per patch-round.
This quantity can in turn be used to estimate the overall logical error rate given the lattice surgery constructions in \sec{lattice_surgery}, and then choose the most compact layout that realizes said target.

Normal surface code patches have a logical error probability that scales inverse exponentially in the patch diameter ($d$), linearly versus round count ($r$), and linearly versus patch count ($n$).
Consequently, the asymptotic scaling of normal patches is $\Theta(r \cdot n \cdot \lambda_0^{-d})$ for some error suppression factor $\lambda_0$.
A simple way to predict this scaling is to focus on the length and number of shortest error paths.
Increasing the patch diameter increases the shortest path length, which is the source of the exponential suppression versus $d$.
Doubling the number of rounds (or patches) doubles the number of shortest paths, which causes linear scaling in $r$ (or $n$).

Yoked surface code patches change the length and number of shortest error paths.
In 1D yoked surface codes, shortest error paths correspond to two shortest error paths in different surface code patches.
This doubles the code distance, but introduces quadratic scaling in $r$, the number of rounds between checks of the yoke.
Furthermore, since a logical failure can occur over any two patches in the code, the number of shortest error paths also scales quadratically in $n$.
Based on this, we expect the scaling of the 1D yoked logical error rate to be $\Theta(r^2 \cdot n^2 \cdot \lambda_1^{-d})$, where $\lambda_1$ denotes the increased error suppression factor obtained from doubling the code distance, which we would expect to be at most $\lambda_0^2$.
Note that the quadratic scaling versus $r$ is only for numbers of rounds up to the yoke check period.
Shortest paths from patches in different check periods don't combine to form a shortest error path in the concatenated code (although they can form an effective measurement error).
Consequently, we expect the scaling versus rounds to be quadratic up to the check period, and then linear afterwards.

A similar argument holds for 2D yoked surface codes, for which the distance quadruples, but now the error scaling becomes quartic versus the number of rounds $r$ between yoke checks, while remaining quadratic in the number of patches $n$.
The quartic scaling versus $r$ is similar: within the support of a weight four logical error, any combination of error paths within those patches can cause the error.
The quadratic scaling versus $n$ is because weight four logical errors correspond to four logical errors landing on the corners of a rectangle within the grid.
There are $\Theta(n^2)$ of these rectangles specified by their endpoints across a diagonal.
Based on this, we expect the scaling of the 2D yoked logical error rate to be $\Theta(r^4 \cdot n^2 \cdot \lambda_2^{-d})$, where $\lambda_2$ denotes the increased error suppression factor obtained from quadrupling the code distance, which we would expect to be at most $\lambda_0^4$.

We emphasize that these are simple path-counting heuristics, where we can empirically fit the different $\lambda$ factors and coefficients.
However, these can of course break down.
For example in 2D, the set of higher-weight inner logical errors leading to failure could scale as $o(r^4)$.
Despite this, we observe good agreement with these heuristic scaling laws.

\subsection{Numerics}
\begin{figure}[htb!]
    \centering
    \resizebox{\linewidth}{!}{
        \includegraphics{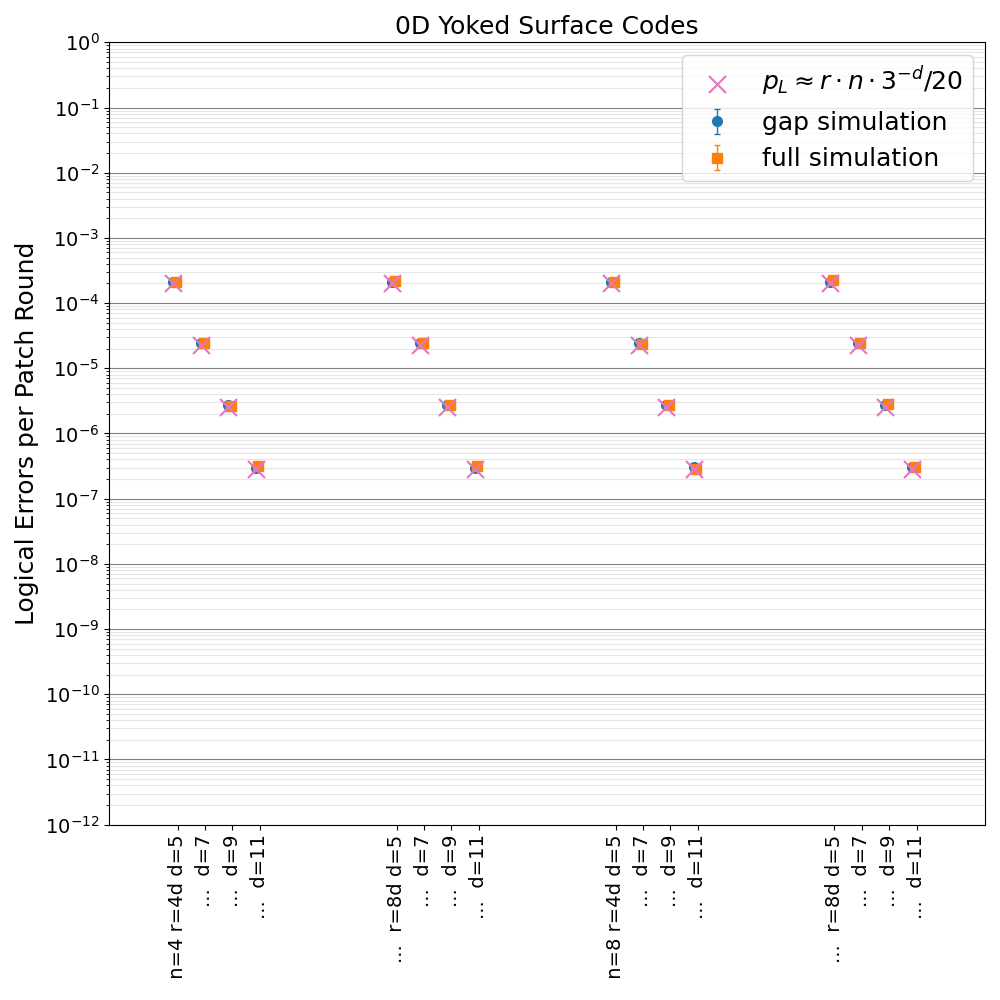}
        \hfill
        \includegraphics{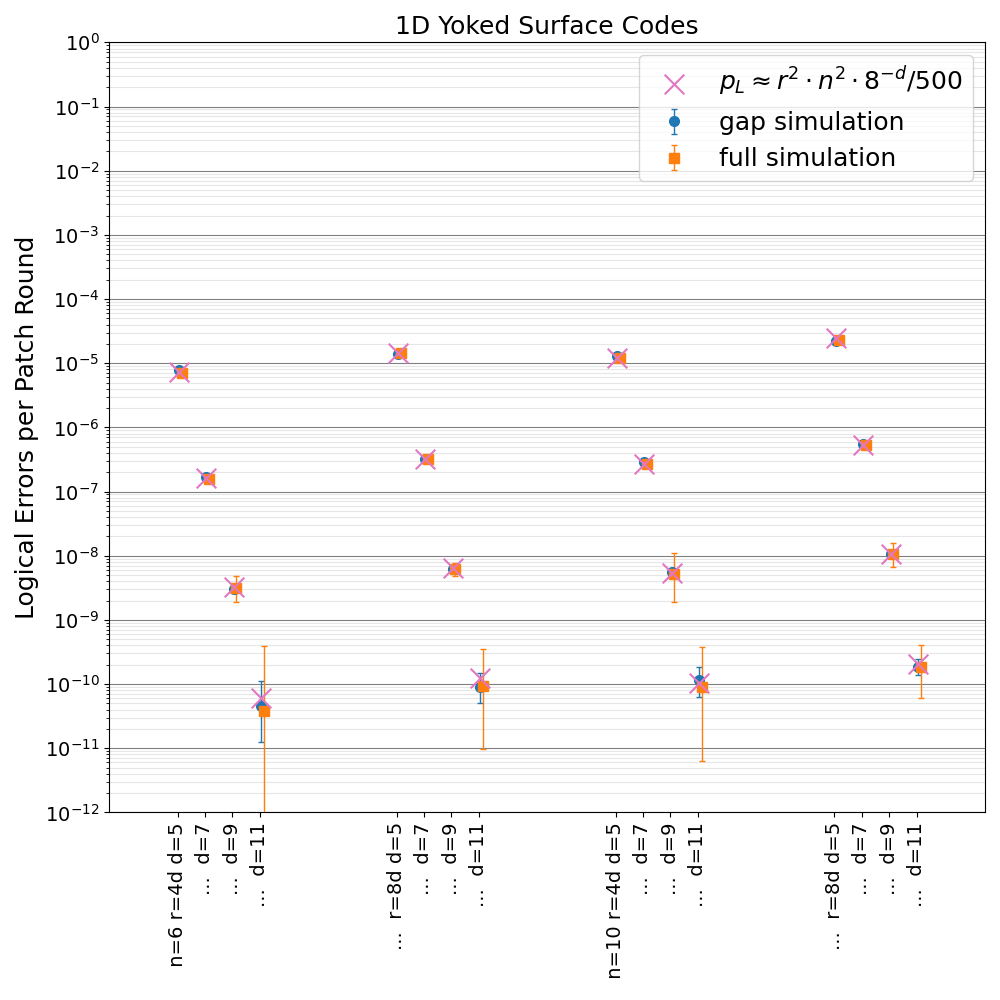}
        \hfill
        \includegraphics{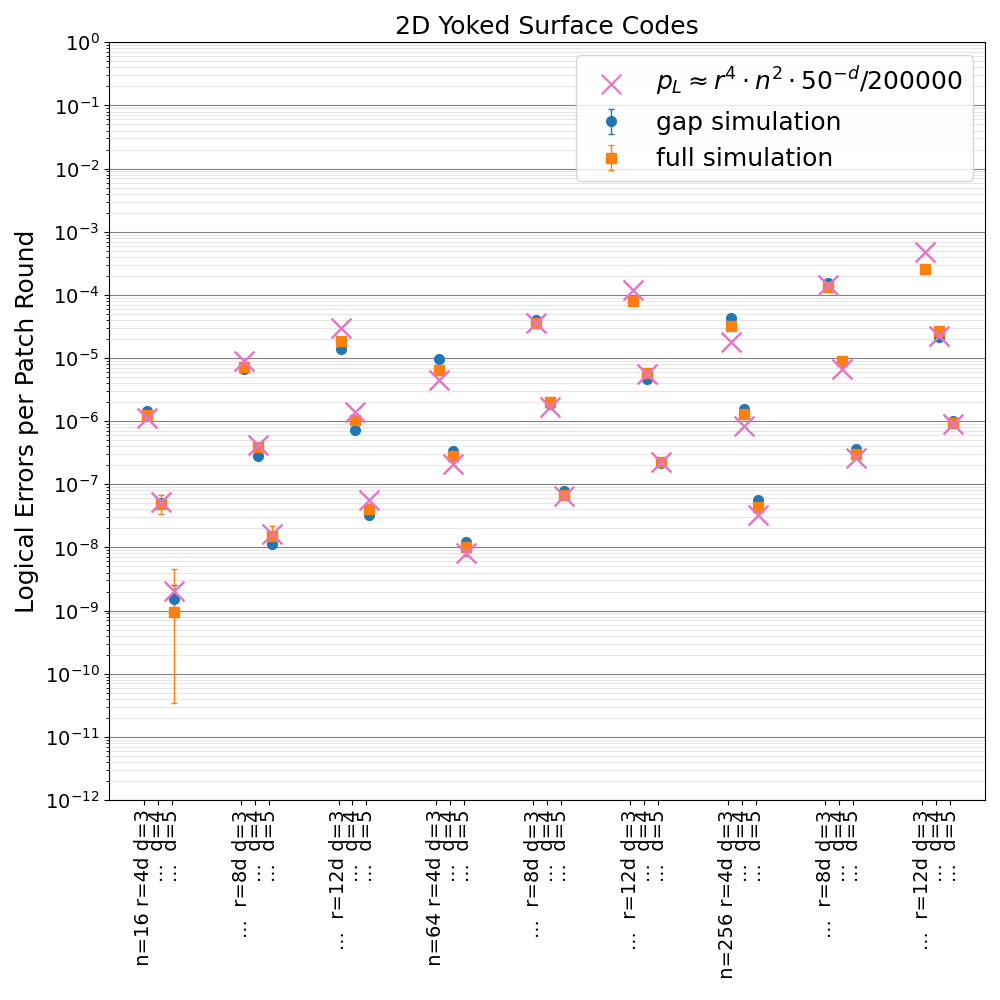}
    }
    \caption{
        A comparison of full simulation versus gap sampling simulation for left: 0D (i.e. normal surface codes), middle: 1D, and right: 2D yoked surface codes. 
        The error rates are reported in terms of logical error per patch-round. Note that this does not account for the overhead introduced by the code itself  or its workspace (i.e in $>0$D, it is somewhat lower than the logical error per logical qubit).
        However, it is convenient for arguing about the total error per outer syndrome cycle.
        We perform simulations over different code block sizes $n$ and different numbers of rounds $r$ in the check.
        We include single-significant-figure fits consistent with the different path-counting scalings that well-approximate the data.
    }
    \label{fig:full_sim}
\end{figure}

We first validate the gap simulation against the full single-outer-round circuit simulation of the inner surface codes - see \fig{full_sim}.
The full simulations were performed by generating stim circuits~\cite{gidney2021stim} to describe the experiments, with yoke detectors added to the boundaries to preserve matchability.
We then decode these experiments using correlated minimum-weight perfect matching.
The circuits that we generate for both generating gap distributions and performing the full simulations use the gateset $\{U_1,CZ,M_Z,R_Z\}$ and the superconducting-inspired noise model SI1000 specified in \app{noise} at an error rate of $10^{-3}$.
We use noiseless time boundaries where the stabilizers and observables are prepared or measured noiselessly.
It's also important that the circuits have many rounds, to reduce distortions from the noiseless time boundaries, as well as minimize the imperfect extrapolation of the gap distribution.
All circuits that we run use gaps extrapolated from at least $10d$ round distributions, where $d$ is the inner surface code patch diameter.

Normally, we would simulate more varied error rates.
The underlying reason for focusing on one error rate is that Monte Carlo simulation of even small yoked surface codes is rather expensive, as the natural size scales span orders of magnitude more qubits and rounds than normal memory experiments.
We were particularly interested in understanding scaling with respect to size, rather than with respect to noise strength, so we sacrificed noise strength diversity in favor of size diversity.

We observe good agreement between the gap simulation and full simulation in $0D$ and $1D$, as well as fairly good agreement in $2D$.
We suspect the small deviations are a result of imperfections in the extrapolation of the gap distribution magnified by the distance-4 outer code.
From these points, we can establish single-significant-figure fits that well-approximate these error rates and are consistent with the path-counting scaling approximations.
These fits are 
\begin{align*}
    p_{L,0} &\approx r_i \cdot n \cdot 3^{-d}/20 \\
    p_{L,1} &\approx r_o \cdot r_i^2 \cdot n^2 \cdot 8^{-d}/500 \\
    p_{L,2} &\approx r_o \cdot r_i^4 \cdot n^2 \cdot 50^{-d}/200000,
\end{align*}
where $p_{L,k}$ indicates the cumulative logical error rate of $k$-D yoked surface codes over $r_o$ rounds of the outer code, with $r_i$ rounds between checks, for size-$n$ code blocks, and distance-$d$ inner surface codes.
Note that, relative to the $3^{-d}$ error suppression factor of the standard surface code, the error suppression factors in 1D (2D) of $8^{-d}$ ($50^{-d}$) fall short of the optimal $9^{-d}$ ($81^{-d}$) projected from doubling (quadrupling) the code distance.

Our final simulations are phenomenological simulations over many rounds. 
In particular, we simulate sizes on the order of the maximum extents of our extrapolations, up to $200d$ rounds between checks and outer blocks of 256 surface code patches.
We simulate 10 rounds of the outer code, with a fixed $100$ $d\times d\times d$ blocks of spacetime contributing to measurement error, which is greater than any of our constructions (see \fig{1d_yoke_pipe_component} and \fig{2d_yoke_pipe_component}).
In the limit of very many $d\times d\times d$ blocks contributing to measurement error, the system should behave like a one-round experiment with $r_o \cdot r_i$ inner rounds as the measurement edges approach weight zero, up to identifying degenerate edges.
However, for the sizes we consider, it appears that the (spacelike) minimum-weight error paths dominate the scaling behavior.

\begin{figure}[htb!]
    \centering
    \resizebox{\linewidth}{!}{
        \includegraphics{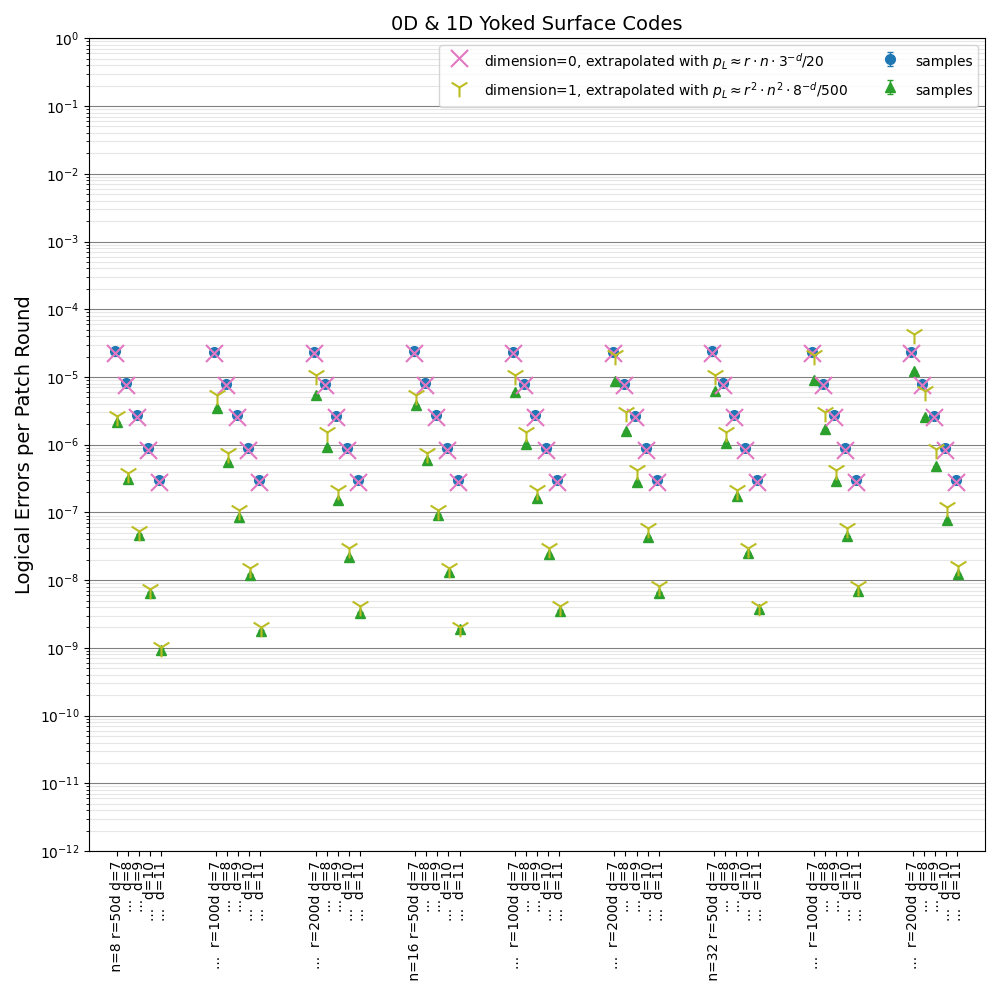}
        \hfill
        \includegraphics{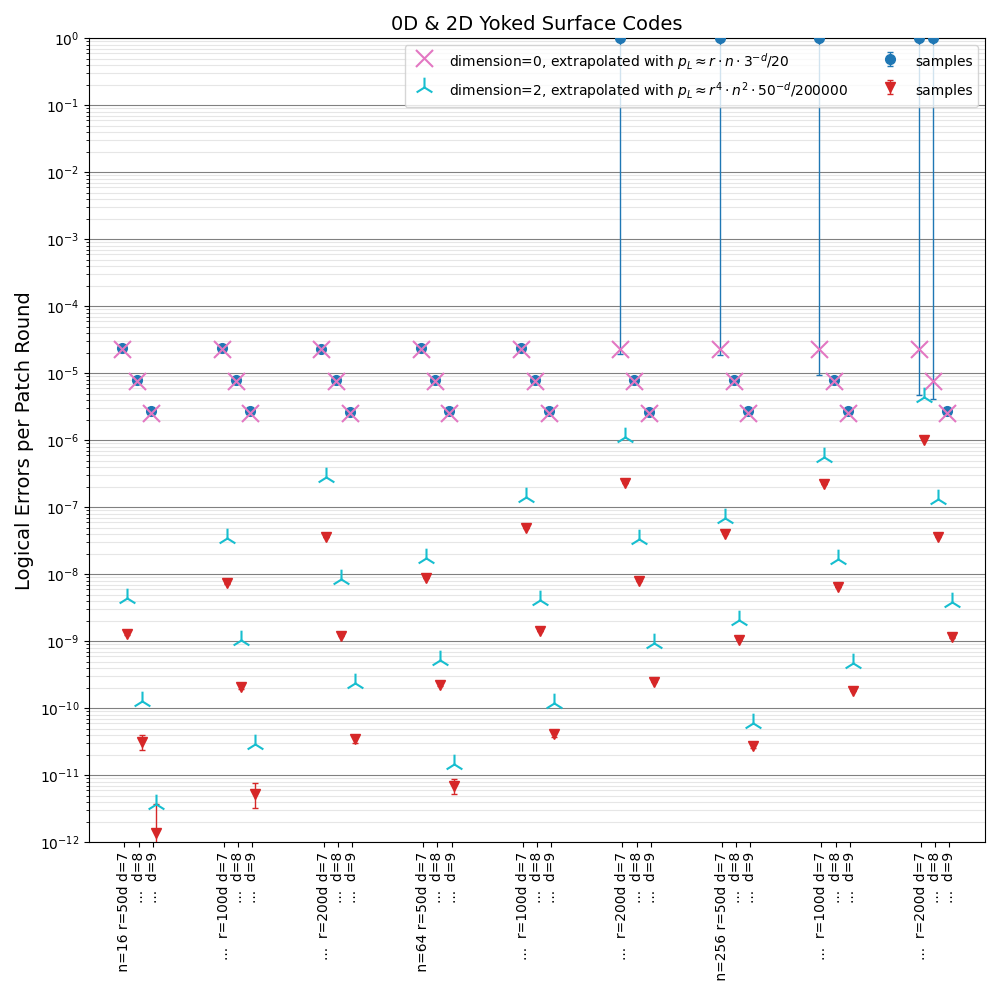}
    }
    \caption{
        Phenomenological simulations of yoked surface codes over 10 rounds of the outer code, with a fixed measurement error rate given by a $100d$ round gap distribution.
        These simulations assume the hook error has been suppressed below the noise floor set by other error mechanisms.
        Left: a comparison of 0D and 1D yoked surface codes with $r$ rounds between checks and length $n$ code blocks.
        Right: a comparison of 0D and 2D yoked surface codes.
        The surface code data points toward the very top right correspond to highly noisy simulations, but with error bars that cover the extrapolations.
    }
    \label{fig:long_sims}
\end{figure}

We present gap simulations of these long phenomenological benchmarks in \fig{long_sims}.
We observe excellent agreement in 0D with the scaling approximations.
For 1D yoked surface codes, we see good agreement, with a slight tendency for the scaling approximations to predict too high a logical error rate.
For 2D yoked surface codes, we see a significant deviation from the scaling approximations predicted from smaller experiments.
There are several potential culprits for this deviation.
One is that it appears the true scaling with number of rounds is not quite $r^4$, and that deviation becomes pronounced in the large round limit.
In fact, were we instead to use two significant figures to estimate the logical error scaling with rounds, even a small change to $r^{3.8}$ would yield good agreement with the observed logical error rates.
Another could be the slightly optimistic extrapolation of the gap distribution to many rounds observed in \fig{gap}.
However, since changing the inner code distance by even one significantly changes the logical error rate, we can obtain a fairly good overhead estimate from extrapolations that are within an order of magnitude.
Furthermore, projecting overheads from scaling laws should provide conservative estimates, as the scaling laws tend to predict higher logical error rates.

\subsection{Footprint estimates}

\begin{figure}[htb!]
    \centering
    \resizebox{\linewidth}{!}{
       \includegraphics{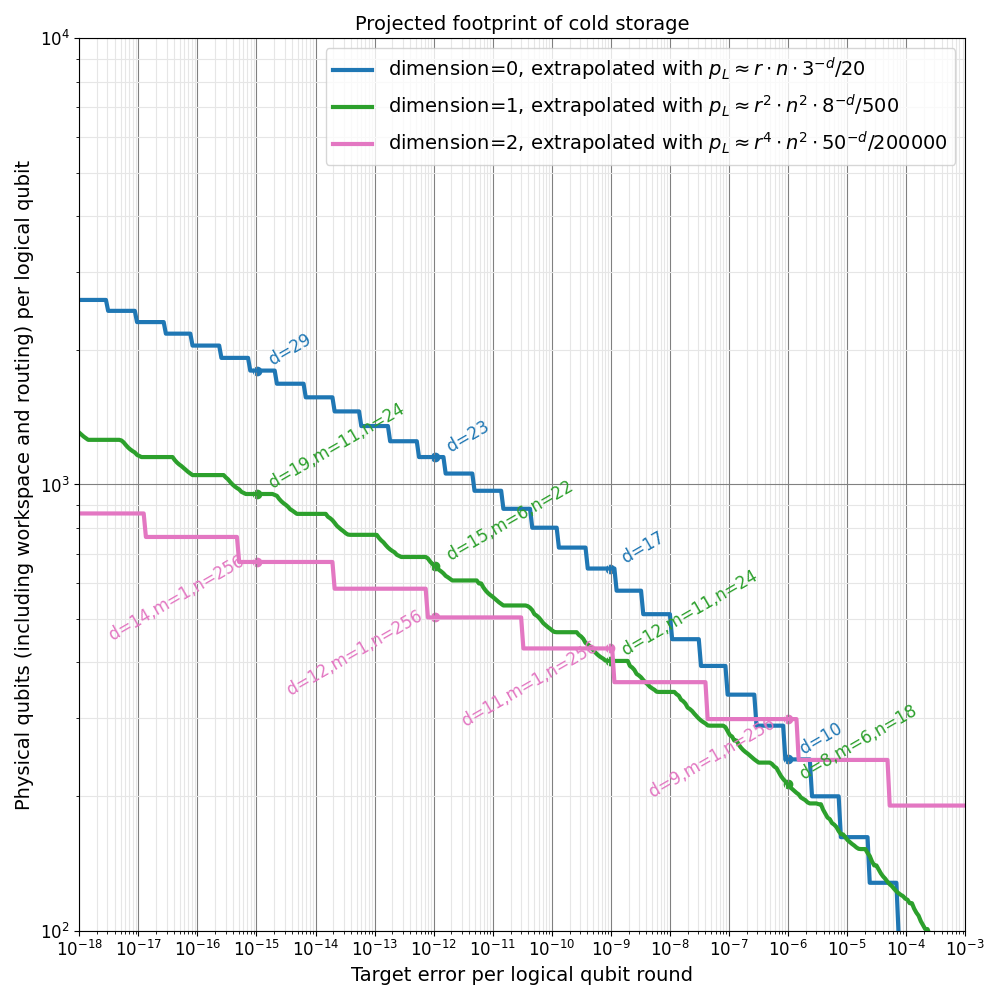}
        \hfill
        \includegraphics{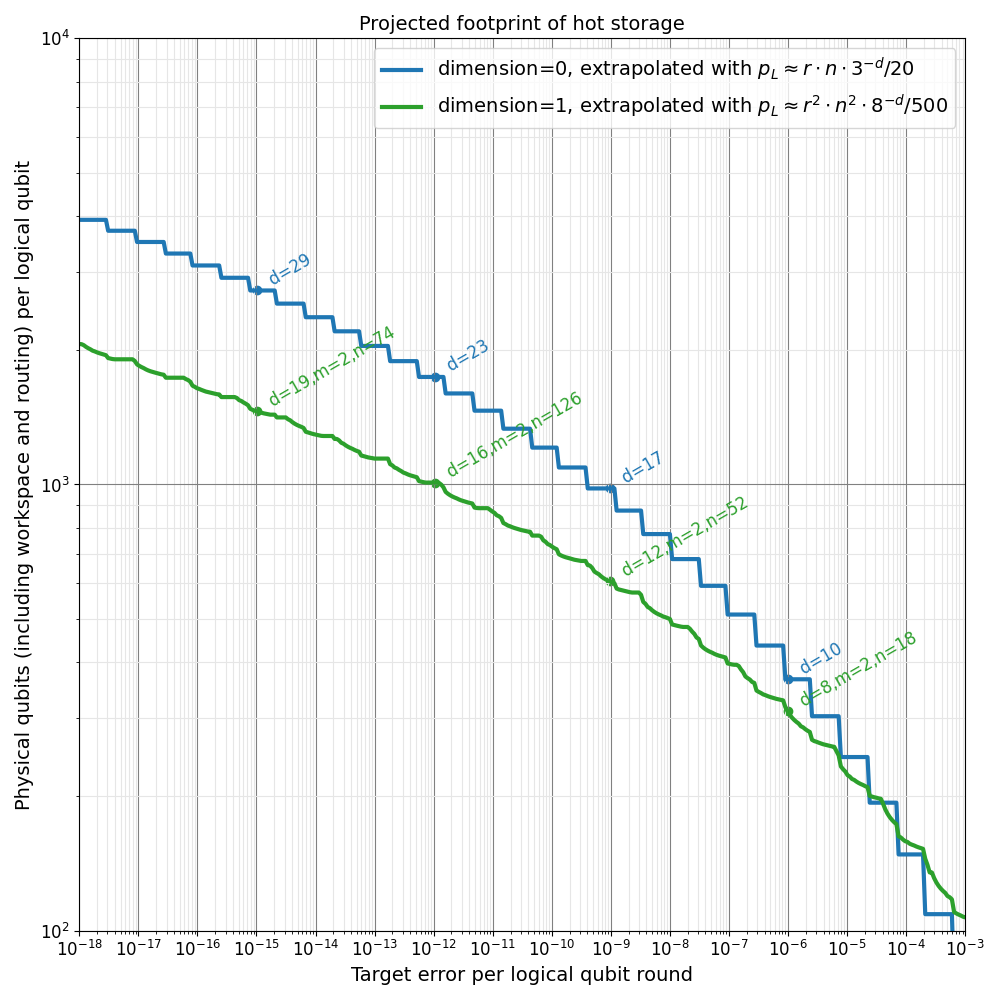}
    }
    \caption{
        Extrapolated footprints of 0D (standard), 1D, and 2D yoked surface codes, including the workspace and access hallway overheads shown in \fig{yoking_hierarchy} and \fig{storage}.
        Projections for cold storage are on the left, and projections for hot storage are on the right.
        For each target logical error rate, various patch diameters $d$, block sizes $n$, and number of blocks $m$ are tried.
        The most efficient layout that meets the target and encodes at most 250 logical qubits is identified using the scaling approximations.
        A patch of diameter $d$ is assumed to cover $2(d+1)^2$ physical qubits to leave some buffer space for lattice surgery.
        The yoked hot storage estimates target an access hallway utilization of 40\%.
    }
    \label{fig:footprint}
\end{figure}

Having given evidence that these heuristic scaling approximations provide plausible, conservative estimates of the logical error rates, we use them to project the footprints required to hit different target logical error rates.
These footprints include the overhead introduced by the workspace required to measure the checks, the overhead introduced by the yokes themselves, as well as the overhead of having an interstitial space between surface codes to mediate the lattice surgery.
We include estimates for cold storage using both 1D and 2D yoked surface codes, as well as hot storage using 1D yoked surface codes - see \fig{footprint}.
Note that we report the target error per round, rather than per the $d$ rounds required to perform a primitive logical operation.
Consequently, the teraquop regime begins around the $10^{-13}-10^{-14}$ logical error rate mark in the figure.

In the teraquop regime, 1D yoked surface codes provide hot storage with nearly twice as many logical qubits per physical qubit as normal surface codes.
2D yoked surface codes provide cold storage with nearly three times as many logical qubits per physical qubit as normal surface codes.
In both cases, as the target logical error rate decreases, the benefits of yoked surface codes become more pronounced.

\section{Conclusion}
\label{sec:conclusion}

In this paper, we described how to ``yoke'' surface codes by measuring row and column logical parity checks along grids of surface codes. 
We estimate that yoked surface codes achieve nearly three times as many logical qubits per physical qubit as standard surface codes when operating in the teraquop regime at a physical error rate of $10^{-3}$.
We also described a hierarchical architecture of hot and cold storage, where we can trade density of encoding for ease of memory access.
In the context of hot storage, we estimate that yoked surface codes achieve nearly twice as many logical qubits per physical qubit as standard surface codes while keeping the logical qubits easily accessible during a computation.
Reducing the quantity of qubits required by surface codes is extremely useful because the required quantity of physical qubits is perhaps the worst aspect of the surface code.
While this is likely not as dramatic as reductions obtainable using LDPC codes with long-range connections, it does not require any additional connectivity and is fairly simple to lay out.

While we have focused on providing simple memory overhead estimates, there are several questions left unanswered.
First, as we enter the large-scale error correction regime, it will be important to develop tooling that makes these types of simulations feasible.
Currently, building the logical stim circuits is a hassle, and Monte Carlo simulations at this scale are difficult to perform exactly.
Building automated tools out of the ZX calculus \cite{bombin2023logical} would be helpful towards performing full-scale simulations of hierarchical memories.
Ultimately, we only provide evidence of these savings through extrapolations - full-scale simulations are needed to verify the actual overhead saved.
Due to the conservative choice of hook error suppression and scaling approximations, we expect that the overheads we report can be improved.

Understanding hierarchical memories in the context of a fault-tolerant circuit is also important.
In focusing on a memory, we have not carefully addressed questions of where and when these memories should be used in a computation, what sort of savings we can expect in that setting, how to gracefully encode, extract, and operate on encoded qubits, and so forth.
These types of questions are important in assessing the overhead of applications of fault-tolerant quantum computers \cite{gidney2021factor}.

Stepping back, in identifying candidate outer codes to concatenate the surface code into, we focused mostly on coding rate and a simple layout.
This is because the inner surface codes are already expensive to employ, and so we focus on high-density parity check codes.
The cost of the surface code pays for itself in its additional features: flexibility to measure larger stabilizers and useful soft information.
In total, we believe that hierarchical memories like the one presented here hold promise for significantly reducing the cost of a surface-code-based fault-tolerant quantum computer.

\section{Contributions}

Peter Brooks did groundwork on using complementary gaps for decoding.
Cody Jones had the idea of concatenating a $Y^{\otimes n}$ stabilizer over the surface code to double the code distance.
Craig Gidney got excited by Cody's idea, did simulations with circuit noise on the inner code and code capacity noise on the outer code, extended the idea to 1D parity check codes, wrote the first draft of the paper, and constructed the 2D parity check codes for an appendix.
Michael Newman got excited by Craig's results, extended the simulations to include gapped phenomenological noise, compiled some of the topological circuits, pushed to focus more on the 2D parity check codes, and rewrote the paper.

\section{Acknowledgements}

We thank Jahan Claes, Austin Fowler, and Matt McEwen for helpful discussions, as well as Noah Shutty for writing the correlated minimum weight perfect matching decoder that we used (derived from PyMatching).
We thank the Google Quantum AI team for creating an environment where this work was possible.

\section{Data availability}

Circuit generation code, stats collected, plotting code, and SketchUp renderings for this paper are available at \cite{gidney_data_2023}.

\printbibliography

@article{de2017zxlattice,
  title={The ZX calculus is a language for surface code lattice surgery},
  author={de Beaudrap, Niel and Horsman, Dominic},
  journal={arXiv preprint arXiv:1704.08670},
  year={2017}
}

@ARTICLE{fowler2012surfacecodereview,
 author={A. G. Fowler and M. Mariantoni and J. M. Martinis and A. N. Cleland},
 title={Surface codes: Towards practical large-scale quantum computation},
 journal={Phys. Rev. A},
 volume={86},
 year={2012},
 pages={032324},
 note={arXiv:1208.0928},
 doi={10.1103/PhysRevA.86.032324},
}

@misc{fowler2018latticesurgery,
  doi = {10.48550/ARXIV.1808.06709},
  url = {https://arxiv.org/abs/1808.06709},
  author = {Fowler,  Austin G. and Gidney,  Craig},
  title = {Low overhead quantum computation using lattice surgery},
  publisher = {arXiv},
  year = {2018},
  copyright = {arXiv.org perpetual,  non-exclusive license}
}

@article{horsman2012latticesurgery,
  title={Surface code quantum computing by lattice surgery},
  author={Horsman, Clare and Fowler, Austin G and Devitt, Simon and Van Meter, Rodney},
  journal={New Journal of Physics},
  volume={14},
  number={12},
  pages={123011},
  year={2012},
  publisher={IOP Publishing},
  doi={10.1088/1367-2630/14/12/123011}
}

@article{litinski2019gameofsurfacecodes,
  doi = {10.22331/q-2019-03-05-128},
  url = {https://doi.org/10.22331/q-2019-03-05-128},
  year = {2019},
  month = mar,
  publisher = {Verein zur Forderung des Open Access Publizierens in den Quantenwissenschaften},
  volume = {3},
  pages = {128},
  author = {Daniel Litinski},
  title = {A Game of Surface Codes: Large-Scale Quantum Computing with Lattice Surgery},
  journal = {Quantum}
}

@inproceedings{shor1994,
  title={Algorithms for quantum computation: Discrete logarithms and factoring},
  author={Shor, Peter W},
  booktitle={Foundations of Computer Science, 1994 Proceedings., 35th Annual Symposium on},
  pages={124--134},
  year={1994},
  organization={Ieee}
}

@article{baspin2023improved,
  title={Improved rate-distance trade-offs for quantum codes with restricted connectivity},
  author={Baspin, Nou{\'e}dyn and Guruswami, Venkatesan and Krishna, Anirudh and Li, Ray},
  journal={arXiv preprint arXiv:2307.03283},
  year={2023}
}

@article{poulin2006optimal,
  title={Optimal and efficient decoding of concatenated quantum block codes},
  author={Poulin, David},
  journal={Physical Review A},
  volume={74},
  number={5},
  pages={052333},
  year={2006},
  publisher={APS}
}

@article{bravyi2023high,
  title={High-threshold and low-overhead fault-tolerant quantum memory},
  author={Bravyi, Sergey and Cross, Andrew W and Gambetta, Jay M and Maslov, Dmitri and Rall, Patrick and Yoder, Theodore J},
  journal={arXiv preprint arXiv:2308.07915},
  year={2023}
}

@article{manes2023distance,
  title={Distance-preserving stabilizer measurements in hypergraph product codes},
  author={Manes, Argyris Giannisis and Claes, Jahan},
  journal={arXiv preprint arXiv:2308.15520},
  year={2023}
}

@article{xu2023constant,
  title={Constant-overhead fault-tolerant quantum computation with reconfigurable atom arrays},
  author={Xu, Qian and Ataides, J and Pattison, Christopher A and Raveendran, Nithin and Bluvstein, Dolev and Wurtz, Jonathan and Vasic, Bane and Lukin, Mikhail D and Jiang, Liang and Zhou, Hengyun},
  journal={arXiv preprint arXiv:2308.08648},
  year={2023}
}

@misc{meister2023efficient,
  title={Efficient soft-decision decoding for hierarchical codes},
  author={Meister, Nadine and Pattison, Christopher A and Preskill, John},
  note={To appear.},
}

@article{gidney2021stim,
  doi = {10.22331/q-2021-07-06-497},
  title = {Stim: a fast stabilizer circuit simulator},
  author = {Gidney, Craig},
  journal = {{Quantum}},
  issn = {2521-327X},
  publisher = {{Verein zur F{\"{o}}rderung des Open Access Publizierens in den Quantenwissenschaften}},
  volume = {5},
  pages = {497},
  month = jul,
  year = {2021}
}

@article{fowler2013optimal,
  title={Optimal complexity correction of correlated errors in the surface code},
  author={Fowler, Austin G},
  journal={arXiv preprint arXiv:1310.0863},
  year={2013},
  doi={10.48550/arXiv.1310.0863}
}

@article{bombin2023logical,
  title={Logical blocks for fault-tolerant topological quantum computation},
  author={Bombin, Hector and Dawson, Chris and Mishmash, Ryan V and Nickerson, Naomi and Pastawski, Fernando and Roberts, Sam},
  journal={PRX Quantum},
  volume={4},
  number={2},
  pages={020303},
  year={2023},
  publisher={APS}
}

@article{gidney2021factor,
  title={How to factor 2048 bit RSA integers in 8 hours using 20 million noisy qubits},
  author={Gidney, Craig and Eker{\aa}, Martin},
  journal={Quantum},
  volume={5},
  pages={433},
  year={2021},
  publisher={Verein zur F{\"o}rderung des Open Access Publizierens in den Quantenwissenschaften},
  doi={10.22331/q-2021-04-15-433},
}

@inproceedings{soeken2020improved,
  title={Improved quantum circuits for elliptic curve discrete logarithms},
  author={Thomas Häner and Samuel Jaques and Michael Naehrig and Martin Roetteler and Mathias Soeken},
  booktitle={Post-Quantum Cryptography: 11th International Conference, PQCrypto 2020, Paris, France, April 15--17, 2020, Proceedings},
  volume={12100},
  pages={425},
  year={2020},
  organization={Springer Nature},
  doi={10.1007/978-3-030-44223-1_23}
}

@article{bravyi2011subsystembacon,
  doi = {10.1103/physreva.83.012320},
  url = {https://doi.org/10.1103/physreva.83.012320},
  year = {2011},
  month = jan,
  publisher = {American Physical Society ({APS})},
  volume = {83},
  number = {1},
  author = {Sergey Bravyi},
  title = {Subsystem codes with spatially local generators},
  journal = {Physical Review A}
}

@misc{gidneyinplaceybasis2023,
  doi = {10.48550/ARXIV.
         2302.07395},
  url = {https://arxiv.org/abs/2302.07395},
  author = {Gidney,  Craig},
  title = {Inplace Access to the Surface Code Y Basis},
  publisher = {arXiv},
  year = {2023},
  copyright = {Creative Commons Attribution 4.0 International}
}

@misc{mcewen2022relaxing,
  doi = {10.48550/ARXIV.2302.02192},
  url = {https://arxiv.org/abs/2302.02192},
  author = {McEwen,  Matt and Bacon,  Dave and Gidney,  Craig},
  title = {Relaxing Hardware Requirements for Surface Code Circuits using Time-dynamics},
  publisher = {arXiv},
  year = {2023},
  copyright = {Creative Commons Attribution 4.0 International}
}

@misc{bombin2022postselect,
  doi = {10.48550/ARXIV.
         2212.00813},
  url = {https://arxiv.org/abs/2212.00813},
  author = {Bombín,  Héctor and Pant,  Mihir and Roberts,  Sam and Seetharam,  Karthik I.},
  title = {Fault-tolerant Post-Selection for Low Overhead Magic State Preparation},
  publisher = {arXiv},
  year = {2022},
  copyright = {arXiv.org perpetual,  non-exclusive license}
}

@misc{litinskyhypercubeshor2023,
  doi = {10.48550/ARXIV.2306.08585},
  url = {https://arxiv.org/abs/2306.08585},
  author = {Litinski,  Daniel},
  title = {How to compute a 256-bit elliptic curve private key with only 50 million Toffoli gates},
  publisher = {arXiv},
  year = {2023},
  copyright = {arXiv.org perpetual,  non-exclusive license}
}

@article{steane1996simpleqec,
  doi = {10.1103/physreva.54.4741},
  url = {https://doi.org/10.1103/physreva.54.4741},
  year = {1996},
  month = dec,
  publisher = {American Physical Society ({APS})},
  volume = {54},
  number = {6},
  pages = {4741--4751},
  author = {A. M. Steane},
  title = {Simple quantum error-correcting codes},
  journal = {Physical Review A}
}

@article{tremblay2022ldpc,
  doi = {10.1103/physrevlett.129.050504},
  url = {https://doi.org/10.1103/physrevlett.129.050504},
  year = {2022},
  month = jul,
  publisher = {American Physical Society ({APS})},
  volume = {129},
  number = {5},
  author = {Maxime A. Tremblay and Nicolas Delfosse and Michael E. Beverland},
  title = {Constant-Overhead Quantum Error Correction with Thin Planar Connectivity},
  journal = {Physical Review Letters}
}

@article{higgott2021hyperbolic,
  doi = {10.1103/physrevx.11.031039},
  url = {https://doi.org/10.1103/physrevx.11.031039},
  year = {2021},
  month = aug,
  publisher = {American Physical Society ({APS})},
  volume = {11},
  number = {3},
  author = {Oscar Higgott and Nikolas P. Breuckmann},
  title = {Subsystem Codes with High Thresholds by Gauge Fixing and Reduced Qubit Overhead},
  journal = {Physical Review X}
}

@misc{pattison2023planarcircuitbounds,
  doi = {10.48550/ARXIV.
         2303.04798},
  url = {https://arxiv.org/abs/2303.04798},
  author = {Pattison,  Christopher A. and Krishna,  Anirudh and Preskill,  John},
  title = {Hierarchical memories: Simulating quantum LDPC codes with local gates},
  publisher = {arXiv},
  year = {2023},
  copyright = {arXiv.org perpetual,  non-exclusive license}
}

@misc{berthusen2023partialmeasure,
  doi = {10.48550/ARXIV.2306.17122},
  url = {https://arxiv.org/abs/2306.17122},
  author = {Berthusen,  Noah and Gottesman,  Daniel},
  title = {Partial Syndrome Measurement for Hypergraph Product Codes},
  publisher = {arXiv},
  year = {2023},
  copyright = {arXiv.org perpetual,  non-exclusive license}
}

@article{hutter2014complementarygap,
  doi = {10.1103/physreva.89.022326},
  url = {https://doi.org/10.1103/physreva.89.022326},
  year = {2014},
  month = feb,
  publisher = {American Physical Society ({APS})},
  volume = {89},
  number = {2},
  author = {Adrian Hutter and James R. Wootton and Daniel Loss},
  title = {Efficient Markov chain Monte Carlo algorithm for the surface code},
  journal = {Physical Review A}
}

@inproceedings{li2020numerical,
  doi = {10.1109/QCE
         49297.2020.00024},
  title={A numerical study of bravyi-bacon-shor and subsystem hypergraph product codes},
  author={Li, Muyuan and Yoder, Theodore J},
  booktitle={2020 IEEE International Conference on Quantum Computing and Engineering (QCE)},
  pages={109--119},
  year={2020},
  organization={IEEE}
}

@article{gottesman2013fault,
  doi = {https://doi.org/10.48550/arXiv.1310.2984},
  title={Fault-tolerant quantum computation with constant overhead},
  author={Gottesman, Daniel},
  journal={arXiv preprint arXiv:1310.2984},
  year={2013}
}

@article{kovalev2013fault,
  doi = {https://doi.org/10.1103/PhysRevA.87.020304},
  title={Fault tolerance of quantum low-density parity check codes with sublinear distance scaling},
  author={Kovalev, Alexey A and Pryadko, Leonid P},
  journal={Physical Review A},
  volume={87},
  number={2},
  pages={020304},
  year={2013},
  publisher={APS}
}

@article{tillich2013quantum,
  doi = {10.1109/TIT.2013.
         2292061},
  title={Quantum LDPC codes with positive rate and minimum distance proportional to the square root of the blocklength},
  author={Tillich, Jean-Pierre and Z{\'e}mor, Gilles},
  journal={IEEE Transactions on Information Theory},
  volume={60},
  number={2},
  pages={1193--1202},
  year={2013},
  publisher={IEEE}
}

@article{panteleev2021degenerate,
  doi = {https://doi.org/10.22331/q-2021-11-22-585},
  title={Degenerate quantum LDPC codes with good finite length performance},
  author={Panteleev, Pavel and Kalachev, Gleb},
  journal={Quantum},
  volume={5},
  pages={585},
  year={2021},
  publisher={Verein zur F{\"o}rderung des Open Access Publizierens in den Quantenwissenschaften}
}

@inproceedings{panteleev2022asymptotically,
  doi = {https://doi.org/10.48550/arXiv.2111.
         03654},
  title={Asymptotically good quantum and locally testable classical LDPC codes},
  author={Panteleev, Pavel and Kalachev, Gleb},
  booktitle={Proceedings of the 54th Annual ACM SIGACT Symposium on Theory of Computing},
  pages={375--388},
  year={2022}
}

@article{baspin2022connectivity,
  doi = {	https://doi.org/10.22331/q-2022-05-13-711},
  title={Connectivity constrains quantum codes},
  author={Baspin, Nou{\'e}dyn and Krishna, Anirudh},
  journal={Quantum},
  volume={6},
  pages={711},
  year={2022},
  publisher={Verein zur F{\"o}rderung des Open Access Publizierens in den Quantenwissenschaften}
}

@article{baspin2022quantifying,
  doi = {https://doi.org/10.1103/PhysRevLett.129.050505},
  title={Quantifying nonlocality: How outperforming local quantum codes is expensive},
  author={Baspin, Nou{\'e}dyn and Krishna, Anirudh},
  journal={Physical Review Letters},
  volume={129},
  number={5},
  pages={050505},
  year={2022},
  publisher={APS}
}

@article{bravyi2010tradeoffs,
  doi = {https://doi.org/10.1103/PhysRevLett.104.050503},
  title={Tradeoffs for reliable quantum information storage in 2D systems},
  author={Bravyi, Sergey and Poulin, David and Terhal, Barbara},
  journal={Physical review letters},
  volume={104},
  number={5},
  pages={050503},
  year={2010},
  publisher={APS}
}

@article{bravyi2009no,
  doi = {10.1088/1367-2630/11/4/043029},
  title={A no-go theorem for a two-dimensional self-correcting quantum memory based on stabilizer codes},
  author={Bravyi, Sergey and Terhal, Barbara},
  journal={New Journal of Physics},
  volume={11},
  number={4},
  pages={043029},
  year={2009},
  publisher={IOP Publishing}
}

@article{breuckmann2017hyperbolic,
  doi = {10.1088/2058-9565/aa7d3b},
  title={Hyperbolic and semi-hyperbolic surface codes for quantum storage},
  author={Breuckmann, Nikolas P and Vuillot, Christophe and Campbell, Earl and Krishna, Anirudh and Terhal, Barbara M},
  journal={Quantum Science and Technology},
  volume={2},
  number={3},
  pages={035007},
  year={2017},
  publisher={IOP Publishing}
}

@article{fawzi2020constant,
  doi = {https://doi.org/10.1145/3434163},
  title={Constant overhead quantum fault tolerance with quantum expander codes},
  author={Fawzi, Omar and Grospellier, Antoine and Leverrier, Anthony},
  journal={Communications of the ACM},
  volume={64},
  number={1},
  pages={106--114},
  year={2020},
  publisher={ACM New York, NY, USA}
}

@misc{gidney_data_2023,
    title = {Data for ``Yoked surface codes''},
    author = {Gidney, Craig and Newman, Michael and Brooks, Peter and Jones, Cody},
    date = {2023-12-06},
    doi = {10.5281/zenodo.10277397},
    url = {https://zenodo.org/record/10277397},
}

\appendix
\clearpage
\section{Noise model}
\label{app:noise}

Simulations in this paper were done using the superconducting-inspired circuit noise model defined in \tab{noise_model}.
The name ``SI1000'' is short for Superconducting Inspired with 1000 nanosecond cycle.

\begin{table}[h]
    \centering
    \begin{tabular}{|r|l|}
    \hline
    Noise channel & Probability distribution of effects
    \\
    \hline
    $\text{MERR}_B(p)$ & $\begin{aligned}
        1-p &\rightarrow M_{B}
        \\
        p &\rightarrow M_{(-1 \cdot B)} \text{\;\;\;\;\;\emph{(i.e. measurement result is inverted)}}
    \end{aligned}$
    \\
    \hline
    $\text{XERR}(p)$ & $\begin{aligned}
        1-p &\rightarrow I
        \\
        p &\rightarrow X
    \end{aligned}$
    \\
    \hline
    $\text{ZERR}(p)$ & $\begin{aligned}
        1-p &\rightarrow I
        \\
        p &\rightarrow Z
    \end{aligned}$
    \\
    \hline
    $\text{DEP1}(p)$ & $\begin{aligned}
        1-p &\rightarrow I
        \\
        p/3 &\rightarrow X
        \\
        p/3 &\rightarrow Y
        \\
        p/3 &\rightarrow Z
    \end{aligned}$
    \\
    \hline
    $\text{DEP2}(p)$ & $\begin{aligned}
        1-p &\rightarrow I \otimes I
        &\;\;
        p/15 &\rightarrow I \otimes X
        &\;\;
        p/15 &\rightarrow I \otimes Y
        &\;\;
        p/15 &\rightarrow I \otimes Z
        \\
        p/15 &\rightarrow X \otimes I
        &\;\;
        p/15 &\rightarrow X \otimes X
        &\;\;
        p/15 &\rightarrow X \otimes Y
        &\;\;
        p/15 &\rightarrow X \otimes Z
        \\
        p/15 &\rightarrow Y \otimes I
        &\;\;
        p/15 &\rightarrow Y \otimes X
        &\;\;
        p/15 &\rightarrow Y \otimes Y
        &\;\;
        p/15 &\rightarrow Y \otimes Z
        \\
        p/15 &\rightarrow Z \otimes I
        &\;\;
        p/15 &\rightarrow Z \otimes X
        &\;\;
        p/15 &\rightarrow Z \otimes Y
        &\;\;
        p/15 &\rightarrow Z \otimes Z
    \end{aligned}$
    \\
    \hline
    \end{tabular}
    \caption{
        Definitions of various noise channels.
        Used by \tab{noise_model}.
    }
    \label{tab:noise_channels}
\end{table}

\begin{table}[h]
    \centering
    \begin{tabular}{|r|l|}
    \hline
    Ideal gate & Noisy replacement
    \\
    \hline
    (any single qubit unitary, including idling) $U_1$ & $\text{DEP1}(p / 10) \cdot U_1$
    \\
    $\text{CZ}$ & $\text{DEP2}(p) \cdot \text{CZ}$
    \\
    \hline
    $R_Z$ & $\text{XERR}(2p) \cdot R_Z$
    \\
    $M_Z$ & $\text{DEP1}(p) \cdot \text{MERR}_Z(5p)$
    \\
    \hline
    (Wait for $M_Z$ or $R_Z$) & $\text{DEP1}(2p)$
    \\
    \hline
    \end{tabular}
    \caption{
        The superconducting-inspired noise model ``SI1000'' used by simulations in this paper.
        The single parameter $p$ sets the two qubit gate error rate, with other error rates defined relative to this rate.
        Measurements are noisiest while single qubit gates are least noisy.
        Qubits that are not reset or measured during layers containing resets or measurements incur additional depolarization on top of other error mechanisms.
        Noise channels are defined in \tab{noise_channels}.
    }
    \label{tab:noise_model}
\end{table}

\clearpage

\section{$Y$-type yokes}
\label{app:y-type}

\begin{figure}[htb!]
    \centering
    \resizebox{\linewidth}{!}{
        \includegraphics{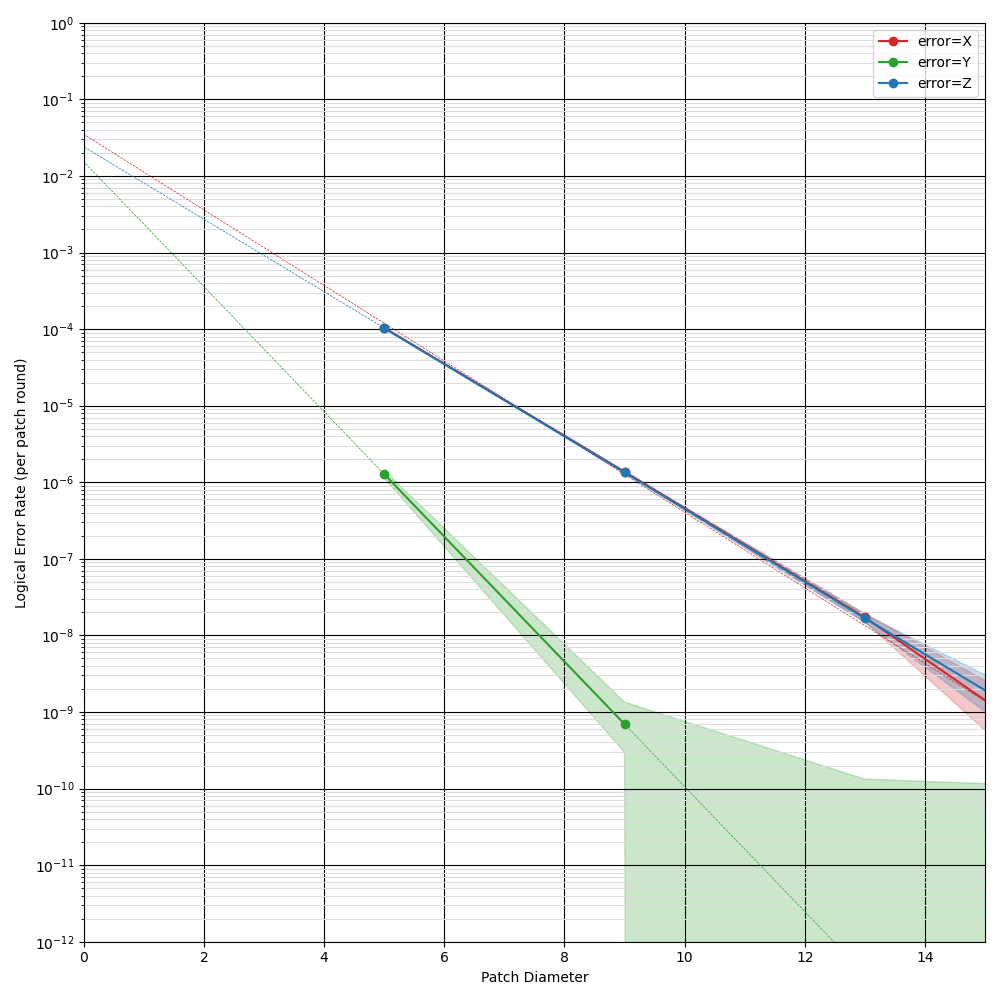}
    }
    \caption{
        Logical X, Y, and Z error rates from a surface code memory circuit.
        Logical errors are biased away from being $Y$-type logical errors.
        Empirically, the $Y$ error rates behave as if the code distance was $1.8 \cdot d$ (vs $1.0 \cdot d$ for $X$ and $Z$ errors), where $d$ is the patch diameter.
    }
    \label{fig:y-bias}
\end{figure}

For 1D yoked surface codes, we could also consider replacing the $X$- and $Z$-type yoke checks with a single $Y$-type yoke check.
The reason is that the inner surface code qubits are highly biased: in a phenomenological noise model on a surface code of distance $d$, the minimum weight of a $Y$-type logical operator is $2d$.
Although in a circuit-level error model the minimum weight of a $Y$-type error is again distance $d$~\cite{gidneyinplaceybasis2023}, the very specific alignment of errors causing this failure makes it relatively rare.
In simulation, for physical error rates around $10^{-3}$, we observe that the surface code behaves like a code with effective distance $1.8d$ against $Y$-type errors - see \fig{y-bias}.

\begin{figure}[htb!]
    \centering
    \resizebox{\linewidth}{!}{
        \includegraphics{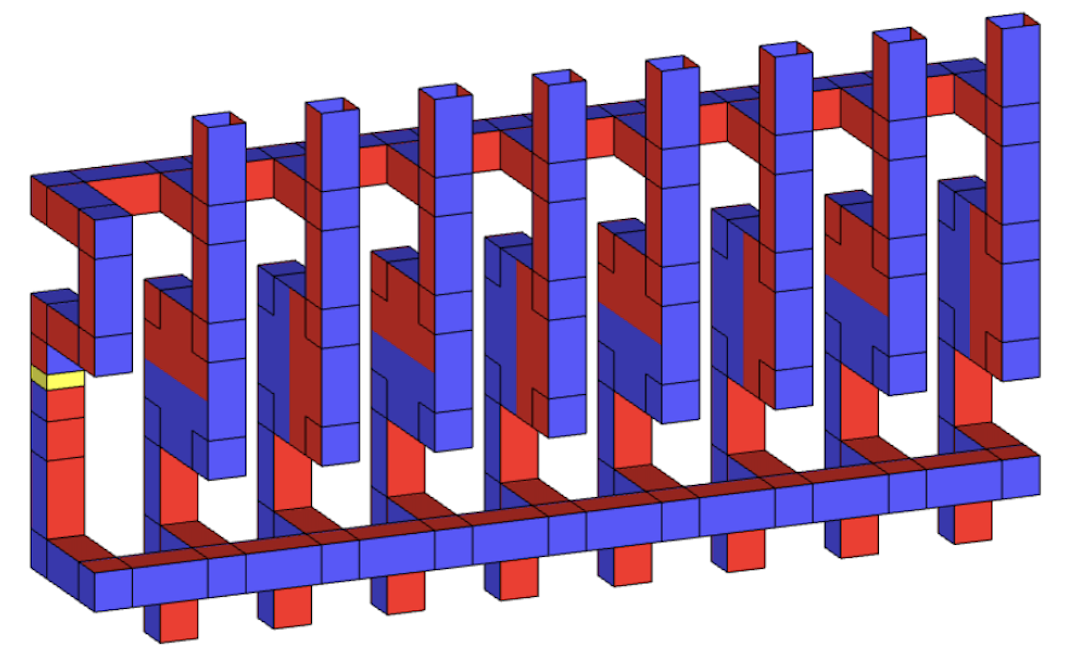}
    }
    \caption{
        Lattice surgery construction for measuring a single $Y$-type check as an alternative 1D yoked surface code, assuming an even block size.
        The yellow block corresponds to a transversal Hadamard operation.
        Beware that this instantiation contains unsuppressed hook errors.
    }
    \label{fig:y-type-check}
\end{figure}

This begs the question: why not use individual $Y$-type checks rather than joint $X$- and $Z$- type checks to increase the number of logical qubits per physical qubit?
There are two reasons - first, while measuring the $Y$-type check, there is a point in the lattice surgery where we may have to increase the size of the surface code to ensure resilience to $Y$-type errors (see \fig{y-type-check}).
Second, the potential savings are reduced by the need to access both the $X$- and $Z$-type boundaries of the surface code in order to perform the $Y$-type check.
Consequently, upon initial consideration, it seems prudent to use standard $X$- and $Z$-type checks.
However, taking advantage of the intrinsic bias of the surface code could prove profitable in future fault-tolerant constructions.

\clearpage

\section{Quantum multi-dimensional parity check codes}
\label{app:mdpc}

Classical multi-dimensional parity check codes (MDPCCs) are defined by arraying bits in an $r$-dimensional tensor. 
They are specified by a list of side lengths $(n_1, \ldots, n_r)$. 
Each parity check is a row check along some dimension, i.e. a rank-$r$ delta tensor $\delta_{i_1, \ldots,\hat{i_\ell},\ldots, i_r}$, where $\hat{i_\ell}$ denotes a missing entry which is the free index - the degree of freedom whose parity is checked.

In 1D, this is simply a single parity check on all the bits. 
In 2D, this corresponds to laying out the bits in an $m \times n$ matrix, with $m$ row parity checks and $n$ column parity checks. 
For example, the $j$th column parity check would be denoted $\delta_{,j}$. 
The total number of bits is $n = \prod_i{n_i}$ and the minimum undetectable error is a configuration of bit flips forming the vertices of an $r$-dimensional rectangle within the array, and so $d=2^r$.

To compute the number of encoded bits $k$, we can count the number of linearly independent parity checks. 
To do this, we iterate over the dimensions - in the $i_1$th direction, we have $n_2n_3\ldots n_r$ parity checks (the area of the hyperface parallel to $i_1$). 
In the $i_2$th direction, we have $(n_1 - 1)n_3n_4 \ldots n_r$ independent parity checks, where $(n_1 - 1)$ accounts for the $n_3n_4\ldots n_r$ parity checks that can be generated from the first $i_1$th direction parity checks and the $(n_1 - 1)n_3n_4 \ldots n_r$ parity checks in the $i_2$th direction. 
Continuing in this way, the $i_k$th direction contributes $(n_1-1)(n_2-1) \ldots (n_{k-1} - 1) n_{k+1} \ldots n_r$ independent parity checks. 
Expanding this polynomial representing $n - k$, we see that $k = \prod\limits_{i=0}^r(n_i - 1)$.

To build quantum MDPCCs, we must construct both $X$-type and $Z$-type checks from this code. 
Unfortunately, simply assigning $X$- and $Z$-type generators from these parity checks directly won't yield a commuting set of stabilizers. 
Instead, when each $n_i \equiv (0 \mod 2^r)$, we can assign $X$-type generators to the parity checks defined by the MDPCC and $Z$-type generators to parity checks defined by a permutation of the MDPCC. 
In particular, because $n_i \equiv (0 \mod 2^r)$, we can write each parity check uniquely as $$\delta_{i_1, \ldots,\hat{i_\ell}, \ldots, i_r} = \delta_{s_1,\ldots,\hat{i_\ell}, \ldots,s_n} \otimes \left(\bigotimes\limits_{k=1}^r \delta_{b_{k_1}, \ldots,\hat{i_\ell}, \ldots, b_{k_r}}\right)$$
where each $\delta_{b_{k_1}, \ldots,\hat{i_\ell}, \ldots, b_{k_r}}$ is a $(2 \times \ldots \times 2)$ rank-$r$ tensor. 
$X$- and $Z$-type stabilizers defined by these parity checks commute if and only if they have even parity when contracted along their ordered indices. 

When two of these $(2\times\ldots\times2)$-tensors differ in any index other than their free index, their contraction is zero. 
If they share the same free index, then their contraction must have even parity since the dimension of each index is two. 
The trouble comes when we have two tensors that are identical in all but their free indices.

The simplest example is any column check $\delta_{,j}$ and row check $\delta_{i,}$ of a matrix - their contraction is one. 
To ensure commutativity, we need to guarantee that every pair of parity checks has a pair of $(2\times\ldots\times2)$ tensors in their decomposition that share a free index in common, so that the overall parity is even. 
We can accomplish this by applying a permutation to the code which cyclically shifts the free indices in each $(2\times\ldots\times2)$ tensor factor. 
Let $\sigma_a$ denote the $(a-1)$-fold cyclic shift of an $r$-element sequence. Then, we apply the permutation:

$$\delta_{s_1,\ldots,\hat{i_\ell}, \ldots,s_n} \otimes \left(\bigotimes\limits_{k=1}^r \delta_{b_{k_1}, \ldots,\hat{i_\ell}, \ldots, b_{k_r}}\right) \mapsto \delta_{s_1,\ldots,\hat{i_\ell}, \ldots,s_n} \otimes \left(\bigotimes\limits_{k=1}^r \delta_{\sigma_k(b_{k_1}, \ldots,\hat{i_{\ell}}, \ldots, b_{k_r})}\right).$$

Note that this does not change the code parameters, but does guarantee that in at least one of the subsystems, the free indices of the $Z$-type and $X$-type checks will be the same. 
Consequently, assigning the $Z$-stabilizers according to the parity checks of this code will yield a commuting set of stabilizers with the same distance. 
In the simple case of $r=2$, this corresponds to transposing the last subsystem.
For $r=3$, this corresponds to cyclically permuting rows to columns to depths in the penultimate subsystem, and rows to depths to columns in the last subsystem.

Because adding the $Z$-stabilizers doubles the number of constraints, we obtain an $[[\prod\limits_{i=1}^r n_i, 2\prod\limits_{i=1}^r(n_i - 1) - \prod\limits_{i=1}^r n_i, 2^r]]$ CSS code family that we call quantum multi-dimensional parity check codes (QPCCs). 
Note that each data qubit participates in two stabilizers per index, and so the total qubit degree is $2r$, while the maximum size stabilizer is $\max_i n_i$. 
In particular, the former constraint yields a matchable code in two dimensions or less.

We gave evidence that concatenating into 1D QPCCs reduced the number of physical qubits required for surface code storage by 1/2, and further by 2/3 when concatenating into 2D QPCCs.
Given this trend, it is natural to ask: is there any promise in continuing to concatenate into higher-dimensional QPCCs? 
They may prove harder to lay out, but by boosting the distance of smaller surface codes with higher distance high-rate outer codes, we can in principle increase the rate of logical qubits per physical qubit. 
However, we eventually hit diminishing returns as the complexity of the logical parity check circuitry increases.

Note also that one of the main advantages of our approach is the high rate of the outer code we concatenate into. 
For any fixed dimension, the rate of a QPCC approaches $1$ as $n \rightarrow \infty$, as the number of constraints scale with the boundary of the array while the encoded degrees of freedom scale with the volume. 
However, lower dimensional codes achieve higher rates at much lower $n$, and in practice, this is an important consideration for integrating these memories into a fault-tolerant computation. 
Restricting to cube-like QPCCs\footnote{This might not always be optimal since, at finite sizes, there could be some maximum qubit limit for which a non-cube-like code is best.} with parameters $[[n^r, 2(n-1)^r - n^r, 2^r]]$, we observe that prohibitively large code sizes are required to achieve e.g. a $75\%$ rate using higher-dimensional codes: for $r=1$, we require $n=8$; for $r=2$, $n=256$; for $r=3$, $n=13824$; and for $r=4$, $n=1048576$. 
This suggests that going beyond $2$D QPCCs might be prohibitively expensive.

\end{document}